\pdfoutput=1
\documentclass{article}

\PassOptionsToPackage{numbers,compress}{natbib}
\usepackage{preprint}

\usepackage[utf8]{inputenc}
\usepackage[T1]{fontenc}
\usepackage[colorlinks=true, linkcolor=blue!70!black, citecolor=blue!70!black, urlcolor=blue!70!black]{hyperref}
\usepackage{url}

\providecommand{\doi}[1]{\href{https://doi.org/#1}{\texttt{doi:#1}}}
\let\oldthebibliography\thebibliography
\renewcommand{\thebibliography}[1]{\oldthebibliography{#1}\renewcommand{\url}[1]{}}
\usepackage{amsmath}
\usepackage{amssymb}
\usepackage{wasysym}
\usepackage{nicefrac}
\usepackage{microtype}
\usepackage{graphicx}
\usepackage{tabularx, booktabs, multirow}
\usepackage{colortbl}
\usepackage{fontawesome}
\usepackage{textcomp}
\usepackage{pifont}
\usepackage{musicography}
\usepackage{xcolor}
\usepackage[shortlabels]{enumitem}
\usepackage{amsthm}

\ifdefined\NewStructureName
  \usepackage{tcolorbox}
\else
  \newenvironment{tcolorbox}[1][]{\par\medskip\noindent\begin{minipage}{\linewidth}\small}{\end{minipage}\medskip\par}
\fi

\definecolor{barcolor}{RGB}{0, 0, 0}
\definecolor{intervalcolor}{RGB}{128, 128, 128}
\definecolor{prcolor}{RGB}{225, 100, 40}
\definecolor{orangehighlight}{RGB}{255, 205, 150}
\definecolor{intervalgray}{RGB}{190, 190, 190}
\definecolor{cyanhighlight}{RGB}{160, 245, 245}
\definecolor{plcolor}{RGB}{40, 148, 148}
\definecolor{noteblack}{RGB}{0, 0, 0}
\definecolor{promptcolor}{RGB}{240, 240, 240}
\definecolor{timestampcolor}{RGB}{233, 229, 252}
\definecolor{velocitycolor}{RGB}{253, 249, 252}

\newcommand{\musiccontext}[1]{
  \par\smallskip
  \begin{quote}
  \noindent\fcolorbox{black!30}{black!8}{
    \parbox{\dimexpr\linewidth-2\fboxsep-2\fboxrule-1.6em}{\small
      \smash{\llap{\raisebox{-1.2ex}{\musFontBig\symbol{71}}\kern0.4em}}
      \textit{Music context.}\; #1}}
  \end{quote}
  \smallskip\par
}

\newcommand{\barline}[1]{{\tiny\ttfamily\colorbox{barcolor}{\textcolor{white}{\textbf{#1}}}}}
\newcommand{\interval}[1]{{\tiny\ttfamily\colorbox{intervalgray}{\textbf{#1}}}}
\newcommand{\token}[1]{{\tiny\ttfamily\fcolorbox{gray!50}{white}{\textbf{#1}}}}
\newcommand{\pr}{{\tiny\ttfamily\colorbox{prcolor}{\textcolor{white}{\textbf{PR:}}}}}
\newcommand{\prpitch}[1]{{\tiny\ttfamily\colorbox{orangehighlight}{\textcolor{noteblack}{\textbf{#1}}}}}
\newcommand{\pl}{{\tiny\ttfamily\colorbox{plcolor}{\textcolor{white}{\textbf{PL:}}}}}
\newcommand{\plpitch}[1]{{\tiny\ttfamily\colorbox{cyanhighlight}{\textcolor{noteblack}{\textbf{#1}}}}}
\newcommand{\prompt}[1]{{\tiny\ttfamily\colorbox{promptcolor}{{\textbf{<|#1|>}}}}}
\newcommand{\stamp}[1]{{\tiny\ttfamily\colorbox{timestampcolor}{{\textbf{<|#1|>}}}}}
\newcommand{\midivel}[1]{{\tiny\ttfamily\colorbox{velocitycolor}{{\textbf{<|#1|>}}}}}

\newcommand{\intermo}{InterMo}

\title{\raisebox{-0.45em}{\includegraphics[height=1.34em]{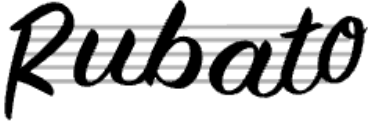}}:  Transcribing Piano Music with Timestamps}

\author{
  Nazif Can Tamer\textsuperscript{$\sharp$} \quad
  Victoria Ebert\textsuperscript{$\sharp$} \quad
  Guang Yang\textsuperscript{$\sharp$} \quad
  Noah A.\ Smith\textsuperscript{$\sharp\flat$} \\[0.5em]
  \textsuperscript{$\sharp$}Paul G. Allen School of Computer Science \& Engineering, University of Washington\\
  \textsuperscript{$\flat$}Allen Institute for AI \\[0.3em]
  \texttt{\{nctamer,nasmith\}@cs.washington.edu}
}

\begin{document}
\maketitle

\begin{abstract}
We consider the conversion of musical recordings into human-readable sheet music annotated with timestamps.  Such output lets a listener clearly visualize \emph{rubato} (temporally expressive playing), a learner diagnose ensemble precision and timing choices against the written music, and a musicology scholar compare performance styles across recordings of the same work.  We introduce (1)  a prompt-conditioned encoder--decoder model, named Rubato, trained to output  (2) a new textual representation for polyphonic music, named InterMo, which we designed for compatibility with sequence-to-sequence training.  Our experiments demonstrate that Rubato produces timestamped piano sheet music from audio with higher notational accuracy than the best existing approaches, which are based on cascades.  We find that even if the cascade is given ground-truth MIDI instead of audio, Rubato performs better, suggesting that the ceiling of existing approaches is primarily representational, not acoustic.  Further, because Rubato is trained on several related tasks (with prompts), it competes with or outperforms the best single-task systems on related but simpler tasks like MIDI note grounding and beat/downbeat detection.  A demo is available at \url{https://nctamer.github.io/rubato-transcription}.
\end{abstract}

\section{Introduction}
\label{sec:intro}

Automatic music transcription has largely been decomposed into specialized tasks such as beat tracking~\citep{foscarin2024beatthis}, audio-to-MIDI\footnote{MIDI (Musical Instrument Digital Interface) is a low-level protocol that encodes pitch as integer semitone numbers with onset times and loudness values. MIDI \emph{transcription} carries no notational structure such as meter, key, or staff assignment.} note detection~\citep{yan2024scoring,edwards2024aria_amt,kong2021high,gardner2021mt3_amt,hawthorne2018maestro}, and MIDI-to-score conversion~\citep{liu2022pm2s_rnn_multi,beyer2024pm2s_transformer,zeng2026bridging}. The resulting components have become strong in isolation, yet accurate intermediate predictions do not always compose into usable sheet music. Errors in the final score may arise at any stage of the pipeline, and gains in intermediate metrics may fail to translate into better final notation for human use in performance, arrangement, learning, or scholarship.

Currently, the dominant approach first transcribes audio into a performance representation that preserves event timing but omits notational structure~\citep{yan2024scoring,edwards2024aria_amt,kong2021high,gardner2021mt3_amt}, then converts this representation into engraved notation through sequence-to-sequence architectures~\citep{beyer2024pm2s_transformer,zeng2026bridging}. This staged decomposition separates temporal grounding from notation. Intermediate representations discard aspects such as staff assignment, metric hierarchy, and pitch spelling that must later be reconstructed downstream, while the resulting score is no longer directly grounded in the recording. Recovering temporal grounding then requires aligning timestamp-only performance data to notation-only score~\citep{peter2023n_asap}---a nontrivial problem in polyphonic music that no prior work has integrated end-to-end.

We address these limitations by introducing \textbf{time-aligned score transcription} as an end-to-end task: Rather than producing a performance representation and reconstructing notation afterwards, the task requires predicting multi-staff polyphonic notation and its temporal alignment jointly. The resulting transcript is both readable as sheet music and directly verifiable against the audio during playback. In this form, a listener can follow \emph{rubato}---where a phrase leans forward, where it lingers, and where expressive freedom reshapes the written pulse---a learner can diagnose timing choices against the score, and a musicology scholar can compare performance styles across recordings of the same work.

\textbf{Rubato} (\autoref{sec:rubato}) is a prompt-conditioned encoder--decoder that transcribes time-aligned scores from audio in a single autoregressive pass. Like multitask automatic speech recognition (ASR) models that transcribe audio into different languages with or without timestamps~\citep{radford2023whisper,hu2025nemo_canary_timestamp}, Rubato uses prompt tokens to select output modes. It can generate time-aligned notation, performed note events, or beat/downbeat annotations in isolation. This is enabled by \textbf{\intermo{}} (\textbf{Inter}vals-and-\textbf{Mo}ments, \autoref{sec:intermo}), a text representation for polyphonic music that serializes sheet music as one-dimensional sequences of pitch-state changes and notated durations. A distinctive feature of \intermo{} is its system of task-specific \emph{dialects}, which expose different aspects of the transcription problem while remaining mutually compatible. These dialects enable multitask training from heterogeneous datasets, where each dataset contributes supervision only for the prompt-conditioned outputs it contains.

Rubato produces lower notational error than all baselines for sheet music transcription, including oracle variants that receive ground-truth MIDI or downbeats as input. We find that transcription quality is limited not only by front-end prediction accuracy, but also by what intermediate representations preserve or discard before downstream score reconstruction. Rubato also achieves state-of-the-art downbeat grounding while remaining competitive with dedicated note-event transcription systems.

\section{Intervals-and-Moments: A Canonical Text Notation for Sheet Music}
\label{sec:intermo}

\begin{figure*}[t]
\centering
\begin{minipage}[t]{0.56\textwidth}
    \vspace{0pt}
    \includegraphics[width=\textwidth]{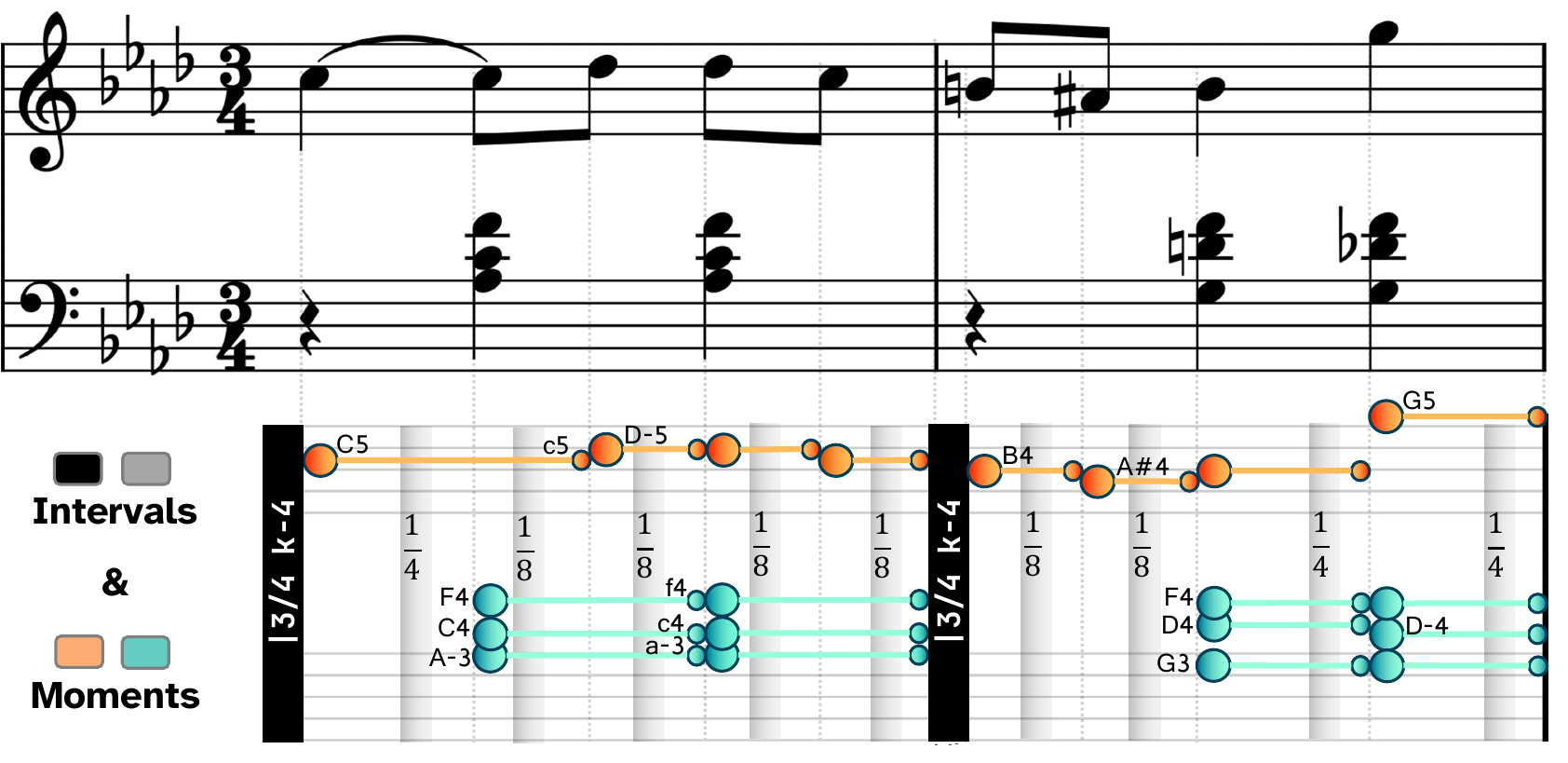}

    \vspace{2pt}
    \setlength{\fboxsep}{1.5pt}
    \setlength{\fboxrule}{0.4pt}
    \newcommand{\z}{\hspace{0pt}}
    \begin{tcolorbox}[colback=gray!8, colframe=gray!20, boxrule=0.5pt, left=0.2em, right=0.2em, top=-0.2em, bottom=-0.2em, fontupper=\scriptsize]
    \setlength{\baselineskip}{7pt}
\textbf{InterMo:}  \barline{|3/4k-4}\z\pr\z\prpitch{C5} \interval{1/4}\z\pl\z\plpitch{A-3}\z\plpitch{C4}\z\plpitch{F4} \interval{1/8}\z\pr\z\prpitch{c5}\z\prpitch{D-5} \interval{1/8}\z\pl\z\plpitch{a-3}\z\plpitch{c4}\z\plpitch{f4}\z\plpitch{A-3}\z\plpitch{C4}\z\plpitch{F4}\z\pr\z\prpitch{d-5}\z\prpitch{D-5} \interval{1/8}\z\prpitch{d-5}\z\prpitch{C5} \interval{1/8}\z\pl\z\plpitch{a-3}\z\plpitch{c4}\z\plpitch{f4}\z\pr\z\prpitch{c5} \barline{|3/4k-4}\z\prpitch{B4} \interval{1/8}\z\prpitch{b4}\z\prpitch{A\#4} \interval{1/8}\z\pl\z\plpitch{G3}\z\plpitch{D4}\z\plpitch{F4}\z\pr\z\prpitch{a\#4}\z\prpitch{B4} \interval{1/4}\z\pl\z\plpitch{g3}\z\plpitch{d4}\z\plpitch{f4}\z\plpitch{G3}\z\plpitch{D-4}\z\plpitch{F4}\z\pr\z\prpitch{b4}\z\prpitch{G5} \interval{1/4}\z\pl\z\plpitch{g3}\z\plpitch{d-4}\z\plpitch{f4}\z\pr\z\prpitch{g5}
    \end{tcolorbox}
\end{minipage}
\hfill
\begin{minipage}[t]{0.42\textwidth}
    \vspace{0pt}
    \setlength{\fboxsep}{1.5pt}
    \setlength{\fboxrule}{0.4pt}
    \tiny

    \fcolorbox{gray}{gray!1}{\parbox{\dimexpr\textwidth-5pt}{\barline{|3/4k-4} New bar, 3/4 time, 4 flats. \pr{} (piano-right) press \prpitch{C5}.}}\\
    \fcolorbox{gray}{gray!1}{\parbox{\dimexpr\textwidth-5pt}{\interval{1/4} Wait a quarter note. \pl{} Press \plpitch{A-3 C4 F4} (sorted by pitch).}}\\
    \fcolorbox{gray}{gray!1}{\parbox{\dimexpr\textwidth-5pt}{\interval{1/8} Wait an eighth note. \pr{} Release \prpitch{c5}, press \prpitch{D-5}.}}\\
    \fcolorbox{gray}{gray!1}{\parbox{\dimexpr\textwidth-5pt}{\interval{1/8} \pl{} Release \plpitch{a-3 c4 f4}, re-press \plpitch{A-3 C4 F4}. \pr{} Release \prpitch{d-5}, re-press \prpitch{D-5}. (staff $\to$ offsets before onsets $\to$ pitch)}}\\
    \fcolorbox{gray}{gray!1}{\parbox{\dimexpr\textwidth-5pt}{\interval{1/8} wait. Release \prpitch{d-5}, press \prpitch{C5} (\pr{} still active).}}\\
    \fcolorbox{gray}{gray!1}{\parbox{\dimexpr\textwidth-5pt}{\interval{1/8} wait. On \pl{} Release \plpitch{a-3 c4 f4}. On \pr{} release \prpitch{c5}.}}\\
    \fcolorbox{gray}{gray!1}{\parbox{\dimexpr\textwidth-5pt}{\barline{|3/4k-4} New bar, 3/4 time, 4 flats.  Press \prpitch{B4}.}}\\
    \fcolorbox{gray}{gray!1}{\parbox{\dimexpr\textwidth-5pt}{\interval{1/8} Release \prpitch{b4}, press \prpitch{A\#4}.}}\\
    \fcolorbox{gray}{gray!1}{\parbox{\dimexpr\textwidth-5pt}{\interval{1/8} \pl{} Press \plpitch{G3 D4 F4}. \pr{} Release \prpitch{a\#4}, press \prpitch{B4}.}}\\
    \fcolorbox{gray}{gray!1}{\parbox{\dimexpr\textwidth-5pt}{\interval{1/4} \pl{} Release \plpitch{g3 d4 f4}, press \plpitch{G3 D-4 F4}. \pr{} Release \prpitch{b4}, press \prpitch{G5}.}}\\
    \fcolorbox{gray}{gray!1}{\parbox{\dimexpr\textwidth-5pt}{\interval{1/4} \pl{} Release \plpitch{g3 d-4 f4}. \pr{} Release \prpitch{g5}. All keys lifted.}}\\
\end{minipage}

\caption{\textbf{\intermo{} representation.} \textbf{Left, top to bottom:} A two-bar piano excerpt, decomposed into metric \emph{intervals} (gray/black) and pitched \emph{moments} (orange = upper staff, teal = lower staff), serialized as a 1D text sequence. \textbf{Right:} English narration of \intermo{}, segmented into interval-piece pretokenization borders which follow a canonical order within score moments (\autoref{subsec:tokenization}). Uppercase = onset, lowercase = offset; \# = sharp, - = flat (e.g., \prpitch{A-3}: A$\flat$3). Staff markers (\pr{}\pl{}, for piano right and left hands, respectively) emitted only on change.}
\label{fig:banner}
\end{figure*}

A piece of sheet music in \intermo{} is built from two primitive types. \textbf{Moments} (orange and teal in \autoref{fig:banner}) mark the pitch state changes (note onsets and offsets) directly discoverable from audio, providing anchors for grounding. \textbf{Intervals} connect the sequence of moments into human-readable notation. \intermo{} contains two types of intervals. Metric intervals segment a measure into discrete moments by encoding the duration between consecutive moments in abstract metrical time (gray in \autoref{fig:banner}; e.g., \interval{1/8}: one eighth-note). Structural intervals, with implied duration of zero, correspond to barlines that segment the transcript into measures.  Musical structure context is present at every measure boundary (e.g., \barline{|3/4k-4}: 3/4 time, 4 flats), resetting tonal context locally such that no measure depends on its predecessors. Interleaving intervals and moments in a fixed order---staves lowest to highest, offsets before onsets sorted by pitch, staff-change markers (\pr{}, \pl{}) emitted only on context switches---yields a canonical text sequence for any score. The right panel of \autoref{fig:banner} narrates each step of this serialization.

\intermo{} has several properties that reduce the ambiguity in autoregressive generation~\citep{vinyals2016order_matters,liu2022symphonynet} and enforce multimodal alignment with audio and sheet music~\citep{jung2025umust}:

\begin{itemize}[leftmargin=*,itemsep=4pt,parsep=0pt,topsep=2pt]
    \item \textbf{Local metric arithmetic.} Intervals specify the metric fraction between consecutive moments, rather than a discrete position (\emph{tick}) accumulated globally (as in MIDI).
    Between consecutive barlines with declared time signature $m/n$, interval values sum to $m/n$ (e.g., \autoref{fig:banner}: \barline{|3/4k-4}\token{=}\interval{1/4}\token{+}\interval{1/8}\token{+}\interval{1/8}\token{+}\interval{1/8}\token{+}\interval{1/8}). Since each measure is metrically self-contained (its time signature is declared at the barline, not inherited from prior context as in printed music), independently decoded audio segments produce metrically valid notation that can be merged at barlines while preserving cross-measure note state, enabling Rubato's parallel chunk decoding (\autoref{sec:rubato}).

    \item \textbf{Bounded open--close matching.} Moments consist of onset/offset events; for each (staff, pitch) pair, these form balanced parentheses (a Dyck-1 word): \prpitch{C5} opens a note, \prpitch{c5} closes it (\autoref{fig:banner}, right). The full pitch state is therefore a $k$-Shuffle-Dyck language~\citep{suzgun2019_k_shuffle_dyck}---definable in C-RASP~\citep{yang2024counting_rasp}, and the formal language for which pre-pretraining yields the largest token-efficiency gains on downstream language tasks~\citep{hu2025prepretrain_formal_lang}. Canonical ordering within each moment (offsets before onsets, sorted by pitch) gives each simultaneous group of events exactly one serialization, resolving the permutation invariance inherent in polyphonic music~\citep{liu2022symphonynet}, so the next-token prediction objective has an unambiguous target at each position.

    \item \textbf{Separable primitives.} Intervals and moments are distinct primitive types that combine into a valid score, yet each follows separate well-formedness constraints: intervals sum to the declared time signature within each measure, and onset/offset events form balanced Dyck-1 words. This separability provides a shared representation space where related tasks naturally correspond to subsets of \intermo{}: MIDI data (without rhythmic abstractions) can be represented as a moment subset with absolute timestamps, while rhythm and meter---including beats and downbeats---can be expressed within the interval subset alone, without pitched events (see \autoref{subsec:dialects}).

    \item \textbf{Semantic unit.} Most music representations treat the note as the inseparable semantic unit~\citep{hsiao2021compound,zeng2021musicbert_octuple,huron1997humdrum}. In \intermo{}, the semantic unit is the consecutive \interval{interval}+\token{moment} pair (each narration step in \autoref{fig:banner}, right): the interval provides rhythmic context, the moment provides pitched events. Timestamps can optionally slot in after each unit, producing \interval{interval}+\token{moment}+\stamp{timestamp} triplets that ground an entire rhythmic cluster with a single acoustic timestamp, embedding the performer's expressive timing into the score notation (\autoref{fig:dialects}; \autoref{subsec:dialects}).
\end{itemize}

\noindent \intermo{} sequences convert to standard engraving formats (Humdrum~\citep{huron1997humdrum}, MEI~\citep{hankinson2011mei}, MusicXML) and render as human-readable notation via Verovio~\citep{pugin2014verovio}.

\section{Rubato: Prompt-Conditioned Transcription with Multitask Training}
\label{sec:rubato}

\begin{figure*}[t!]
\setlength{\lightrulewidth}{0.1pt}
    \centering
      \makebox[\linewidth][c]{\includegraphics[width=\linewidth]{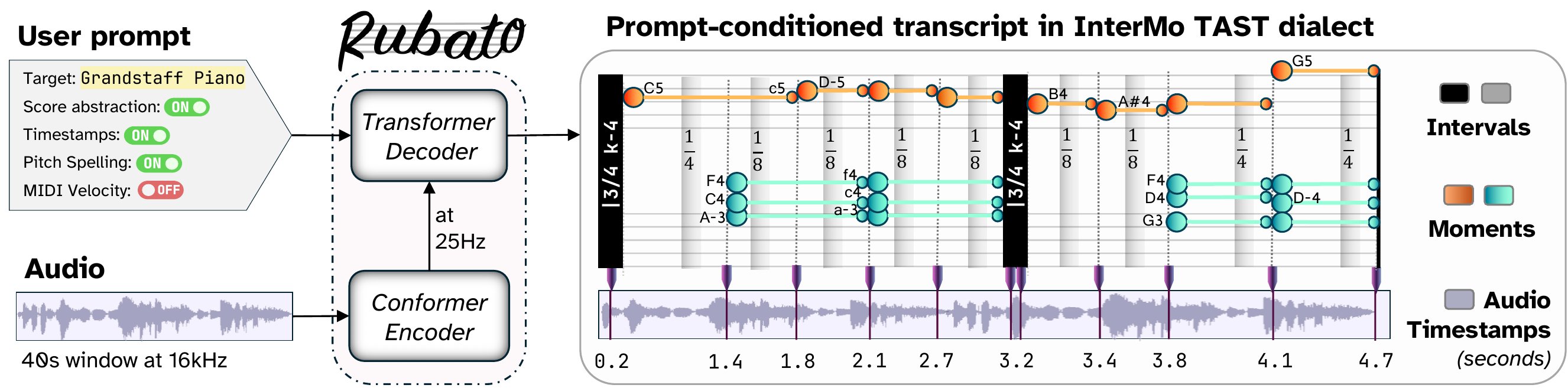}}

    \vspace{4pt}
    \noindent\makebox[\textwidth][c]{\scalebox{0.75}{
    {\small
    \begin{tabular}{l@{\hspace{0.1cm}}r@{\hspace{0.1cm}}l@{\hspace{0.1cm}}p{11cm}}
    \addlinespace[5pt]
    \toprule
    \textbf{Dialect} & \textbf{Prompt}  & $\rightarrow$  & \textbf{Target example (first bar of figure above)} \\
    \midrule

\textit{\underline{TAST}} & \prompt{piano}\prompt{score}\prompt{timestamp}\prompt{spell} & $\rightarrow$ & \barline{|3/4k-4}\pr{}\prpitch{C5}\stamp{0.20} \interval{1/4}\pl{}\plpitch{A-3 C4 F4}\stamp{1.40} \interval{1/8}\pr{}\prpitch{c5 D-5}\stamp{1.80} \interval{1/8}\pl{}\plpitch{a-3 c4 f4}\plpitch{A-3 C4 F4}\pr{}\prpitch{d-5}\prpitch{D-5}\stamp{2.10} \interval{1/8}\prpitch{d-5}\prpitch{C5}\stamp{2.70} \interval{1/8}\pl{}\plpitch{A-3 C4 F4}\pr{}\prpitch{c5}\stamp{3.20} \barline{|3/4k-4}\prpitch{B4}\stamp{3.20}\ldots\\

    \addlinespace[5pt]

    \textit{TAST$_{\textit{lite}}$} & \prompt{piano}\prompt{score}\prompt{timestamp} & $\rightarrow$ & \barline{|3/4k-4}\pr{}\prpitch{C5}\stamp{0.20} \interval{1/4}\pl{}\plpitch{G\#3 C4 F4}\stamp{1.40} \interval{1/8}\pr{}\prpitch{c5 C\#5}\stamp{1.80} \interval{1/8}\pl{}\plpitch{g\#3 c4 f4}\plpitch{G\#3 C4 F4}\pr{}\prpitch{c\#5}\prpitch{C\#5}\stamp{2.10} \interval{1/8}\prpitch{c\#5}\prpitch{C5}\stamp{2.70} \interval{1/8}\pl{}\plpitch{G\#3 C4 F4}\pr{}\prpitch{c5}\stamp{3.20} \barline{|3/4k-4}\prpitch{B4}\stamp{3.20}\ldots\\

    \addlinespace[3pt]
    \multicolumn{4}{l}{\itshape \textbf{T}ime \textbf{A}ligned \textbf{S}core \textbf{T}ranscription: full score notation aligned to audio through timestamps (TAST); without notational pitch spelling (TAST$_{\textit{lite}}$).}\\
    \midrule

    \textit{A2S} & \prompt{piano}\prompt{score}\prompt{spell} & $\rightarrow$ &
    \barline{|3/4k-4}\pr{}\prpitch{C5} \interval{1/4}\pl{}\plpitch{A-3 C4 F4} \interval{1/8}\pr{}\prpitch{c5 D-5} \interval{1/8}\pl{}\plpitch{a-3 c4 f4}\plpitch{A-3 C4 F4} \pr{}\prpitch{d-5}\prpitch{D-5} \interval{1/8}\prpitch{d-5}\prpitch{C5} \interval{1/8}\pl{}\plpitch{a-3 c4 f4}\pr{}\prpitch{c5} \barline{|3/4k-4}\prpitch{B4}\ldots\\
    \addlinespace[5pt]

    \textit{A2S$_{\textit{lite}}$} & \prompt{piano}\prompt{score} & $\rightarrow$ &
    \barline{|3/4k-4}\pr{}\prpitch{C5} \interval{1/4}\pl{}\plpitch{G\#3 C4 F4} \interval{1/8}\pr{}\prpitch{c5 C\#5} \interval{1/8}\pl{}\plpitch{g\#3 c4 f4}\plpitch{G\#3 C4 F4} \pr{}\prpitch{c\#5}\prpitch{C\#5} \interval{1/8}\prpitch{c\#5}\prpitch{C5} \interval{1/8}\pl{}\plpitch{g\#3 c4 f4}\pr{}\prpitch{c5} \barline{|3/4k-4}\prpitch{B4}\ldots\\

    \addlinespace[3pt]
    \multicolumn{4}{l}{\itshape \textbf{A}udio-\textbf{t}o-\textbf{S}core Transcription: score notation matching \autoref{sec:intermo} (A2S), same without pitch spelling (A2S$_{\textit{lite}}$)---Training only.}\\
    \midrule
    \addlinespace[5pt]

    \textit{AMT$_{\textit{lite}}$} & \prompt{piano}\prompt{timestamp} & $\rightarrow$ &
    \prpitch{C5}\stamp{0.20} \plpitch{G\#3}\stamp{1.37} \plpitch{F4}\stamp{1.40} \plpitch{C4}\stamp{1.41}
    \prpitch{C\#5}\stamp{1.80} \prpitch{c5}\stamp{1.81}
    \plpitch{g\#3} \plpitch{f4}\stamp{2.00} \prpitch{c\#5}\stamp{2.01} \plpitch{c4}\stamp{2.02}
    \plpitch{G\#3}\stamp{2.08} \prpitch{C\#5}\stamp{2.10} \plpitch{C4} \plpitch{F4}\stamp{2.11}
    \prpitch{C5}\stamp{2.70} \prpitch{c\#5}\stamp{2.72}
    \prpitch{c5}\stamp{3.16} \plpitch{g\#3} \plpitch{f4}\stamp{3.17} \plpitch{c4}\stamp{3.18}\ldots\\
    \addlinespace[5pt]
    \textit{\underline{AMT}} & \prompt{piano}\prompt{timestamp}\prompt{MIDI} & $\rightarrow$ &
    \midivel{CC64:on}\stamp{0.10}
    \prpitch{C5}\midivel{vel:46}\stamp{0.20} \plpitch{G\#3}\midivel{vel:49}\stamp{1.37} \plpitch{F4}\midivel{vel:43}\stamp{1.40} \plpitch{C4}\midivel{vel:56}\stamp{1.41}
    \prpitch{c5}\stamp{1.61} \midivel{CC64:off}\stamp{1.70} \prpitch{C\#5}\midivel{vel:46}\stamp{1.80}
    \plpitch{g\#3} \plpitch{f4}\stamp{2.00} \prpitch{c\#5}\stamp{2.01} \plpitch{c4}\stamp{2.02}
    \plpitch{G\#3}\midivel{vel:36}\stamp{2.08} \prpitch{C\#5}\midivel{vel:44}\stamp{2.10} \plpitch{F4}\midivel{vel:46} \plpitch{C4}\midivel{vel:39}\stamp{2.11}
    \prpitch{C5}\midivel{vel:43}\stamp{2.70} \prpitch{c\#5}\stamp{2.72}
    \prpitch{c5}\stamp{3.16} \plpitch{g\#3} \plpitch{f4}\stamp{3.17} \plpitch{c4}\stamp{3.18}\ldots\\
    \addlinespace[3pt]
    \multicolumn{4}{l}{\itshape \textbf{A}utomatic \textbf{M}IDI \textbf{T}ranscription: pitched events with timestamp grounding (AMT$_{\textit{lite}}$); with MIDI-specific "velocity" and pedal controls (AMT).}\\
    \midrule
    \addlinespace[5pt]
        \textit{DBD$_{\textit{plus}}$} & \prompt{beats}\prompt{score}\prompt{timestamp} & $\rightarrow$ &
    \barline{|3/4k-4}\stamp{0.20} \interval{1/4} \token{*}\stamp{1.40} \interval{1/4} \token{*}\stamp{2.10} \interval{1/4} \barline{|3/4k-4}\stamp{3.20}\ldots\\
    \addlinespace[5pt]
    \textit{\underline{DBD}} & \prompt{beats}\prompt{timestamp} & $\rightarrow$ &
    \barline{|}\stamp{0.20} \token{*}\stamp{1.40} \token{*}\stamp{2.10} \barline{|}\stamp{3.20}\ldots\\
    \addlinespace[5pt]

    \multicolumn{4}{l}{\itshape \textbf{D}ownbeat and \textbf{B}eat \textbf{D}etection: rhythmic anchors in the metrical scaffold (DBD$_{\textit{plus}}$); beat (\token{*}) and downbeats (\barline{|}) without meter and key (DBD).}\\
    \midrule
    \addlinespace[5pt]

    \end{tabular}}}}

    \caption{Top: Rubato model is trained to transcribe the same musical passage into different \intermo{} dialects conditioned on user prompt. Bottom: Eight dialects (tasks) used in training. Dialects used at inference (\underline{\textit{TAST, AMT, DBD}}) are underlined; the remaining five are training-only, exposing the model to different facets of the same content. Color coding is a reading aid and does not correspond to actual tokens, which are learned through UnigramLM ``Interval-Piece'' tokenization (see \autoref{subsec:tokenization}).
    }
    \label{fig:dialects}
\vspace*{-\baselineskip}
\end{figure*}

Rubato is a prompt-conditioned encoder--decoder using the architecture of Canary-180M-Flash~\citep{puvvada2024canary,hu2025nemo_canary_timestamp}, trained from scratch: a Fast Conformer encoder paired with a Transformer decoder (${\approx}$180M parameters; \autoref{fig:dialects}, top). Its primary task is time-aligned score transcription (TAST): producing sheet music anchored in the audio timeline. Training it end-to-end requires learning score notation, timestamp grounding, and the alignment between the two---aspects that existing datasets cover unevenly in quality and scale. Existing datasets are each suited to learning different aspects: timestamp grounding from real audio (MAESTRO~\citep{hawthorne2018maestro}), score notation from large-scale score collections (PDMX~\citep{long2025pdmx}), or aligned scores with beat annotations at limited scale ((n)ASAP~\citep{foscarin2020asap,peter2023n_asap}). We bridge these sources through a dialect system that projects \intermo{} into task-specific subsets --- each matching the annotations a particular dataset provides --- and multitask training.

\subsection{Dialects and Prompt Conditioning}
\label{subsec:dialects}

Each dialect in \autoref{fig:dialects} is a projection of the full \intermo{} representation, selected by prompt tokens. The richest dialect, \textbf{TAST} (Time-Aligned Score Transcription) produces the complete output: score notation with pitch spelling, grounded in performer time through timestamps. Pitch spelling (\prompt{spell} prompt) distinguishes enharmonic equivalents---A$\sharp$4, B$\flat$4, and C\musDoubleFlat{}5 are the same key on a piano but encode different harmonic functions---analogous to resolving homophones in speech transcription. By removing one or more components from TAST, we arrive at other dialects, allowing each dataset to contribute training signal at the level of annotation it provides (\autoref{tab:training_mix}).

Removing \prompt{timestamp}s yields \textbf{A2S} (Audio-to-Score), trainable at scale on PDMX's large score collection with synthesized audio (without fine-grained temporal alignment). Removing \prompt{spell}ing yields the \textit{lite} variants, which share pitch vocabulary with MIDI but retain score structure. Removing \prompt{score} structure entirely yields the \textbf{AMT} (Automatic MIDI Transcription) dialects, retaining only moment primitives with timestamps---the level at which MAESTRO, with its precise real-audio timestamps but no score annotations, can provide training signal. AMT$_\textit{lite}$ produces non-quantized pitch events with timing, while AMT adds 129 MIDI-specific vocabulary items for velocity (\midivel{vel:N}) and sustain pedal (\midivel{CC64:on/off}), making its output MIDI-like~\citep{oore2020midilike_token} and closer to what a Disklavier records. The \textbf{DBD} (Downbeat and Beat Detection) dialects remove pitched events, retaining interval primitives with timestamps and an additional beat marker token (\token{*}), trained on human beat annotations from (n)ASAP.

\subsection{Interval-Piece Tokenization}
\label{subsec:tokenization}

The tokenizer determines how the model segments its output; in autoregressive models, tokenization choices directly affect learning efficiency~\citep{bostrom2020bpe_suboptimal}. We use SentencePiece UnigramLM~\citep{kudo2018unigramLM} on \intermo{} pretokenized at interval boundaries---the points where score time advances---and fit on A2S and A2S$_\textit{lite}$ text, with timestamps and prompt tokens added as predefined vocabulary. The learned merges are the musical analogue of subwords in speech and language tokenizers: merges of the symbolic surface humans read on a score (\autoref{fig:banner})---pitch labels, staff markers \pr{}/\pl{}, interval fractions, articulations. Frequent merges include bigrams (\token{b-5 A-5}), metric fragments (\interval{1/16}\pl\plpitch{e3 F\#3}), chord-pieces (\token{F3 A3 D4}), and notes with articulations (\token{B-3.}: staccato, \token{C2> C3>}: accent). Rare combinations decompose into well-attested components (\token{D\#}~+~\token{8}, or \token{7/}~+~\token{12}), with robustness to such decompositions learned through subword regularization.

The 8,000-token vocabulary used across dialects (\autoref{fig:dialects}) has two parts. The ${\sim}$3{,}570 \emph{semantic} tokens are the learned interval pieces above. The remaining \emph{non-semantic} tokens are control and grounding metadata that carry no linguistic structure and are held out of UnigramLM merge learning: ${\sim}$4{,}000 timestamps at 10\,ms granularity (\stamp{0.00}--\stamp{40.00}), 129 MIDI tokens (\midivel{vel:N}, \midivel{CC64:on/off}), 1 beat marker \token{*}, ${\sim}$40 prompt tokens, and 256 byte-fallback tokens for lossless encoding (rarely invoked, since \intermo{} uses a closed ASCII alphabet). A sheet music sequence and its timestamped variant share the same interval-piece token boundaries; timestamps simply insert between them without restructuring the sequence. 
\subsection{Training and Inference}
\label{subsec:training}

Our \textbf{training mix} is summarized in \autoref{tab:training_mix}. \textbf{MAESTRO~v3}~\citep{hawthorne2018maestro} provides 159 hours of real piano audio with Disklavier MIDI timestamps accurate to ${\sim}$3ms; it feeds only the AMT dialects. \textbf{(n)ASAP}~\citep{foscarin2020asap,peter2023n_asap} enriches a subset of MAESTRO recordings with beat, downbeat, key/time-signature, and score annotations; its 30 hours, segmented with overlapping windows, feed all dialect families except AMT.\footnote{ASAP's audio-MIDI pairs are a subset of MAESTRO, hence omitted from AMT to prevent duplicates.}
Because ASAP is a major source of data contamination in the score transcription literature, we adopt a conservative split that holds out more recordings than the standard partition, and evaluate on two disjoint test sets (\autoref{sec:experiments}). \textbf{PDMX}~\citep{long2025pdmx} is a large-scale collection of public-domain scores from MuseScore with no associated audio. We extract grand-staff piano parts and segment into 4--32 measure utterances (average ${\sim}$20\,s, up to 40\,s). We synthesize audio via DawDreamer~\citep{braun2021dawdreamer} in two variants: non-expressive (constant tempo) and expressive (with timing deviations and occasional note mismatches generated by VirtuosoNet~\citep{jeong2019virtuosonet}). The expressive renderings teach the model to recover the underlying score despite performance deviations in timing and dynamics. Each utterance is synthesized with one of eight piano VST instruments, each with two room/microphone configurations, yielding sixteen timbral variants for augmentation. We release the score excerpts and synthesized utterances for reproducibility.

\begin{table}[t]
\definecolor{zebracol}{gray}{0.965}
\definecolor{sectioncol}{gray}{0.88}
\centering
\small
\caption{Training data sources and dialect (task) coverage. Hours are raw audio without augmentation; utterance counts (in thousands, combining two variants per task family) include data augmentation. Synthetic audio is only used in training and all tests are solely conducted on real recordings.}
\label{tab:training_mix}
\setlength{\tabcolsep}{4pt}
\begin{tabular}{llrrrrr}
\toprule
\textbf{Source} & \textbf{Audio} & \textbf{Hours} & \textbf{TAST} & \textbf{A2S} & \textbf{AMT} & \textbf{DBD} \\
\midrule
MAESTRO~\citep{hawthorne2018maestro} & real & 159 & --- & --- & 804k & --- \\
\rowcolor{zebracol}
(n)ASAP~\citep{foscarin2020asap,peter2023n_asap} & real & 30 & 214k & 214k & --- & 465k \\
PDMX~\citep{long2025pdmx} & synthesized & 2{,}071 & 511k & 1{,}002k & 436k & 521k \\
\bottomrule
\end{tabular}
\end{table}

Training a single model across tasks with very different sequence lengths requires several adaptations:

\begin{itemize}[leftmargin=*,itemsep=4pt,parsep=0pt,topsep=2pt]
\item\textbf{Subword regularization}~\citep{kudo2018unigramLM}. We use ($\alpha{=}0.25$) to stochastically vary segmentation during training, so components of rare pitch and duration combinations receive gradient signal even when their merged-token forms appear only a few times.

\item\textbf{Token weighting}~\citep{clark2026molmo2}. Multi-task sequences vary enormously in length: an AMT sequence for a 40-second excerpt can be two orders of magnitude longer than a DBD sequence for the same audio. We normalize the cross-entropy loss by $1/\sqrt{|T|}$, where $|T|$ is the output sequence length, preventing long-sequence tasks from dominating gradient updates.

\item \textbf{Timestamp label smoothing.} Music grounding tasks require finer temporal precision than the encoder's 40\,ms frame rate. We therefore predict decoder timestamps at 10\,ms resolution. To handle the mismatch between encoder frame rate and decoder granularity, we apply ordinal label smoothing selectively to timestamp tokens: the correct bin receives weight $P_{\text{center}} = 0.9$, and the remaining mass is distributed over a quadratically decaying window of $\pm 5$ bins (110\,ms total span):
\begin{equation*}
q(i)=
\begin{cases}
P_{\text{center}} & i = y \\[2pt]
\textstyle\frac{1-P_{\text{center}}}{Z_y}(w+1-|i-y|)^2 & i \in \mathcal{N}_y \\[4pt]
0 & \text{otherwise}
\end{cases}
\end{equation*}
where $w=5$, $\mathcal{N}_y=\{\,i:\ 0<|i-y|\leq w\,\}$ is the neighborhood of the target bin $y$, and $Z_y$ normalizes the distribution to unit mass.

\item\textbf{Discrete tiling.} For utterances shorter than 40\,s, timestamps concentrate in a fraction of the 4{,}000-bin range, leaving most bins unvisited during training. We uniformly sample audio start offsets so that every timestamp position appears during training.

\end{itemize}

\textbf{Inference.} Long-form audio is decoded in 40\,s encoder windows with 50\% hop. Timestamp tokens beyond 20\,s act as end-of-sequence signals, so the decoder terminates naturally at the 20\,s boundary without decoding the full window. The extra 20\,s of encoder context provides right-side acoustic look-ahead at each chunk boundary. Since sequence length varies considerably across tasks, inference speed is prompt-dependent: the real-time multiplier (RTFx, measured on a single NVIDIA L40S with beam width 4) is ${\sim}$9$\times$ for AMT, ${\sim}$21$\times$ for TAST, and ${\sim}$112$\times$ for DBD, reflecting how subword tokenization of score notation compresses sequences relative to unstructured MIDI output.

\section{Experiments}
\label{sec:experiments}

We evaluate Rubato on three held-out test sets of real piano recordings. \textbf{ASAP} (102 recordings) follows the standard MAESTRO test split~\citep{liu2022pm2s_rnn_multi}. \textbf{ASAP-Beyer} (25 recordings) is a competing split defined by~\citet{beyer2024pm2s_transformer}. \textbf{ATEPP}~\citep{zhang2022atepp} is a large collection of YouTube piano recordings, a subset of which include paired scores.\footnote{ATEPP also provides automatically transcribed MIDI via~\citet{kong2021high}; our evaluation uses only the paired scores.} From this subset, we drop every recording whose score is sourced from ASAP to avoid cross-dataset memorization; the filtered test set contains 1{,}495 recordings across 106 scores. Because ASAP is used in training all baselines with varying splits,\footnote{M2ST is trained on 98/102 scores from ASAP; ASAP-Beyer appears in the training data used by all AMT front-ends; Rubato training data excludes both ASAP and ASAP-Beyer test recordings.} ATEPP provides the fairest comparison across models: it is the largest set unseen by all models, and includes substantial variability in recording conditions, including historical performances.

\subsection{Sheet Music Transcription}
\label{sec:score_transcription}

\definecolor{contamcol}{rgb}{0.68, 0.48, 0.48}
\definecolor{zebracol}{gray}{0.965}
\providecommand{\contam}[1]{\textcolor{contamcol}{#1}}

\begin{table*}[t]
\centering
\caption{Sheet music quality: OMR-NED ($\downarrow$, \%),  $\pm$ bootstrap 95\% CI (10k resamples) over pieces where the pipeline produced an evaluable score. $n_{\text{fail}}$ counts missing outputs (cascade/conversion failures or blank Gemini responses). \contam{Faded cells ($^{*}$)} mark training-data overlap: on ASAP, M2ST is trained on 98/102 test pieces; on ASAP-Beyer, 21/25 test performances are in the MAESTRO training split, seen by every AMT front-end and by PM2S. Best per column \textbf{bold}, second best \underline{underlined}.}
\label{tab:notation_quality}
\vspace{7pt}
\small
\setlength{\tabcolsep}{4pt}
\begin{tabular}{p{1.8cm} l rc rc rc }
\toprule
 &  & \multicolumn{2}{c}{ATEPP ($n{=}1495$)} & \multicolumn{2}{c}{ASAP ($n{=}102$)} & \multicolumn{2}{c}{ASAP-Beyer ($n{=}25$)} \\
\cmidrule(lr){3-4} \cmidrule(lr){5-6} \cmidrule(lr){7-8}
 & System & OMR-NED & $n_{\text{fail}}$ & OMR-NED & $n_{\text{fail}}$ & OMR-NED & $n_{\text{fail}}$ \\
\midrule
\multirow{4}{1.8cm}{\textit{End-to-End}}
 & \textbf{Rubato (TAST)} & \textbf{75.9$_{\scriptscriptstyle\pm0.9}$} &  & \textbf{64.3$_{\scriptscriptstyle\pm3.9}$} &  & \textbf{78.7$_{\scriptscriptstyle\pm5.0}$} &  \\
 & \cellcolor{zebracol}Gemini 3.1 Pro & \cellcolor{zebracol}--- & \cellcolor{zebracol} & \cellcolor{zebracol}98.6$_{\scriptscriptstyle\pm0.3}$ & \cellcolor{zebracol}12 & \cellcolor{zebracol}98.9$_{\scriptscriptstyle\pm0.4}$ & \cellcolor{zebracol}2 \\
 & \hspace{2em} + \textit{in-Context Learning} & --- &  & 97.6$_{\scriptscriptstyle\pm0.4}$ &  & 97.9$_{\scriptscriptstyle\pm0.6}$ &  \\
 & \cellcolor{zebracol}Gemini 3.1 Pro (\textit{Reasoning}) & \cellcolor{zebracol}--- & \cellcolor{zebracol} & \cellcolor{zebracol}98.6$_{\scriptscriptstyle\pm0.4}$ & \cellcolor{zebracol} & \cellcolor{zebracol}98.3$_{\scriptscriptstyle\pm0.8}$ & \cellcolor{zebracol} \\
\midrule
\multirow{1}{1.8cm}{\textit{Cascade (db.)}}
 & Beat-This $\to$ Piano-A2S & 88.9$_{\scriptscriptstyle\pm0.3}$ &  & 86.6$_{\scriptscriptstyle\pm1.2}$ &  & 89.7$_{\scriptscriptstyle\pm2.0}$ &  \\
\arrayrulecolor{gray!50}\specialrule{0.3pt}{2pt}{2pt}\arrayrulecolor{black}
\multirow{8}{1.8cm}{\textit{Cascade (MIDI)}}
 & \cellcolor{zebracol}Tkun $\to$ M2ST & \cellcolor{zebracol}\underline{85.2$_{\scriptscriptstyle\pm0.5}$} & \cellcolor{zebracol}1 & \cellcolor{zebracol}\contam{$^{*}$\underline{69.1$_{\scriptscriptstyle\pm3.9}$}} & \cellcolor{zebracol} & \cellcolor{zebracol}\contam{$^{*}$89.3$_{\scriptscriptstyle\pm2.1}$} & \cellcolor{zebracol} \\
 & Bytedance $\to$ M2ST & 86.5$_{\scriptscriptstyle\pm0.4}$ &  & \contam{$^{*}$76.5$_{\scriptscriptstyle\pm3.2}$} &  & \contam{$^{*}$90.3$_{\scriptscriptstyle\pm1.8}$} &  \\
 & \cellcolor{zebracol}Aria $\to$ M2ST & \cellcolor{zebracol}85.6$_{\scriptscriptstyle\pm0.4}$ & \cellcolor{zebracol} & \cellcolor{zebracol}\contam{$^{*}$76.6$_{\scriptscriptstyle\pm3.2}$} & \cellcolor{zebracol} & \cellcolor{zebracol}\contam{$^{*}$\underline{86.8$_{\scriptscriptstyle\pm2.9}$}} & \cellcolor{zebracol} \\
 & MT3 $\to$ M2ST & 88.8$_{\scriptscriptstyle\pm0.3}$ &  & \contam{$^{*}$86.0$_{\scriptscriptstyle\pm1.3}$} &  & \contam{$^{*}$90.6$_{\scriptscriptstyle\pm1.7}$} &  \\
 & \cellcolor{zebracol}Tkun $\to$ PM2S & \cellcolor{zebracol}92.0$_{\scriptscriptstyle\pm0.2}$ & \cellcolor{zebracol}8 & \cellcolor{zebracol}89.5$_{\scriptscriptstyle\pm1.1}$ & \cellcolor{zebracol}2 & \cellcolor{zebracol}\contam{$^{*}$92.7$_{\scriptscriptstyle\pm1.6}$} & \cellcolor{zebracol}\contam{2} \\
 & Bytedance $\to$ PM2S & 92.5$_{\scriptscriptstyle\pm0.2}$ & 1 & 90.8$_{\scriptscriptstyle\pm1.1}$ &  & \contam{$^{*}$93.8$_{\scriptscriptstyle\pm1.3}$} &  \\
 & \cellcolor{zebracol}Aria $\to$ PM2S & \cellcolor{zebracol}93.2$_{\scriptscriptstyle\pm0.2}$ & \cellcolor{zebracol}3 & \cellcolor{zebracol}90.8$_{\scriptscriptstyle\pm1.2}$ & \cellcolor{zebracol} & \cellcolor{zebracol}\contam{$^{*}$94.1$_{\scriptscriptstyle\pm1.3}$} & \cellcolor{zebracol} \\
 & MT3 $\to$ PM2S & 93.3$_{\scriptscriptstyle\pm0.2}$ & 135 & 92.2$_{\scriptscriptstyle\pm0.7}$ &  & \contam{$^{*}$93.1$_{\scriptscriptstyle\pm1.8}$} &  \\
\midrule
\multirow{1}{1.8cm}{\textit{Oracle (db.)}}
 & \cellcolor{zebracol}Oracle Db. $\to$ Piano-A2S & \cellcolor{zebracol}--- & \cellcolor{zebracol} & \cellcolor{zebracol}77.8$_{\scriptscriptstyle\pm2.5}$ & \cellcolor{zebracol} & \cellcolor{zebracol}87.7$_{\scriptscriptstyle\pm2.8}$ & \cellcolor{zebracol} \\
\arrayrulecolor{gray!50}\specialrule{0.3pt}{2pt}{2pt}\arrayrulecolor{black}
\multirow{2}{1.8cm}{\textit{Oracle (MIDI)}}
 & Oracle MIDI $\to$ M2ST & --- &  & \contam{$^{*}$69.3$_{\scriptscriptstyle\pm3.6}$} &  & 87.9$_{\scriptscriptstyle\pm3.0}$ &  \\
 & \cellcolor{zebracol}Oracle MIDI $\to$ PM2S & \cellcolor{zebracol}--- & \cellcolor{zebracol} & \cellcolor{zebracol}89.8$_{\scriptscriptstyle\pm1.1}$ & \cellcolor{zebracol}1 & \cellcolor{zebracol}\contam{$^{*}$92.6$_{\scriptscriptstyle\pm2.1}$} & \cellcolor{zebracol}\contam{1} \\
\bottomrule
\end{tabular}
\end{table*}
We evaluate the notation quality of Rubato's TAST output using OMR-NED~\citep{martinez2025omr_ned}, the normalized edit distance over engraved sheet music primitives (such as noteheads, beams, accidentals, barlines, time signatures---the elements musicians read).
OMR-NED directly measures the human edit effort needed to correct the transcript into the reference sheet music.

Since no prior work produces sheet music from audio without staged decomposition, we compare with \textbf{cascade baselines} that operate through intermediate representations (downbeats or MIDI). \textbf{Piano-A2S}~\citep{zeng2024a2s_piano_short} is an audio-to-score system with custom hierarchical decoders that processes 5-bar chunks and requires downbeat positions for chunking; we use Beat-This~\citep{foscarin2024beatthis} as its front-end, and evaluate with oracle downbeats where available. \textbf{PM2S}~\citep{liu2022pm2s_rnn_multi} converts performance MIDI to a quantized \emph{score MIDI} through separate RNN heads for beats, key, time signature, note value, and staff. \textbf{M2ST}~\citep{beyer2024pm2s_transformer} is a performance MIDI to score Transformer that emits compound tokens whose attributes are predicted by independent feedforward heads~\citep{hsiao2021compound}, producing a purely symbolic output with no temporal grounding. Notably, M2ST is trained on a closed-source MuseScore corpus; the data scale comparison therefore favors this cascade baseline. We pair PM2S and M2ST with standard AMT front-ends (enumerated in \autoref{sec:related}); where available, we also evaluate oracle variants using ground-truth Disklavier MIDI. As a non-specialized reference, we include \textbf{Gemini~3.1~Pro}, evaluated zero-shot and with in-context learning (\autoref{app:gemini_prompts}).

From real piano recordings, Rubato (prompted for TAST) produces sheet music with lower OMR-NED than every cascade baseline on all test sets (\autoref{tab:notation_quality}). On ASAP, where oracle inputs are available, replacing predicted MIDI with oracle Disklavier MIDI does not change OMR-NED for either MIDI cascade (within confidence intervals): suggesting that, within this decomposition, downstream structure recovery—not audio prediction—is the dominant source of error. Replacing predicted downbeats with oracle downbeats, in contrast, improves Piano-A2S by 8.8 OMR-NED points (86.6 → 77.8), indicating that recovering measure boundaries from expressive performance is itself a substantial part of the audio-to-score problem.
Rubato outperforms both oracle variants on ASAP, demonstrating that an end-to-end formulation can avoid these intermediate failure modes.

\definecolor{zebracol}{gray}{0.965}
\begin{table}[t]
\centering
\caption{Timestamp grounding accuracy: F1 (\%, $\uparrow$) $\pm$ bootstrap 95\% CI (10k resamples). Models producing sheet music with timestamps are evaluated on both beat and note grounding; the cascade baseline uses Tkun, the best-performing MIDI transcriber for downstream tasks. Downbeat/Beat Detection and MIDI Note Detection are single-task baselines. $^{\dagger}$Not directly predicted; approximated post hoc from the transcribed score. TAST does not predict velocity (---); it encodes related information through notational abstractions such as accents. Best per column \textbf{bold}, second best \underline{underlined}.}
\label{tab:grounding}
\small
\setlength{\tabcolsep}{4pt}
\begin{tabular}{p{3.5cm} lccccc}
\toprule
 &  & \multicolumn{3}{c}{ASAP} & \multicolumn{2}{c}{MAESTRO} \\
\cmidrule(lr){3-5} \cmidrule(lr){6-7}
 & Model & F1$_{\text{downbeat}}$ & F1$_{\text{beat}}$ & F1$_{\text{note}}$ & F1$_{\text{note}}$ & F1$_{\text{note+vel}}$ \\
\midrule
\multirow{2}{3.5cm}{\textit{Timestamp Accuracy of Sheet Music Transcription}}
 & \textbf{Rubato (TAST)} & \textbf{67.8$_{\scriptscriptstyle\pm3.1}$} & \textcolor{gray}{75.8$_{\scriptscriptstyle\pm2.9}$$^{\dagger}$} & 91.0$_{\scriptscriptstyle\pm1.8}$ & 87.1$_{\scriptscriptstyle\pm2.0}$ & --- \\
 & \cellcolor{zebracol}Tkun $\to$ PM2S & \cellcolor{zebracol}22.4$_{\scriptscriptstyle\pm3.7}$ & \cellcolor{zebracol}56.0$_{\scriptscriptstyle\pm1.8}$ & \cellcolor{zebracol}95.2$_{\scriptscriptstyle\pm0.9}$ & \cellcolor{zebracol}92.3$_{\scriptscriptstyle\pm0.8}$ & \cellcolor{zebracol}91.8$_{\scriptscriptstyle\pm0.9}$ \\
\midrule
\multirow{2}{3.5cm}{\textit{Beat Detection}}
 & \textbf{Rubato (DBD)} & \underline{65.2$_{\scriptscriptstyle\pm2.7}$} & \textbf{82.6$_{\scriptscriptstyle\pm2.5}$} & --- & --- & --- \\
 & \cellcolor{zebracol}Beat-This & \cellcolor{zebracol}64.9$_{\scriptscriptstyle\pm2.5}$ & \cellcolor{zebracol}\underline{79.9$_{\scriptscriptstyle\pm2.6}$} & \cellcolor{zebracol}--- & \cellcolor{zebracol}--- & \cellcolor{zebracol}--- \\
\midrule
\multirow{5}{3.5cm}{\textit{MIDI Note Detection (AMT)}}
 & \textbf{Rubato (AMT)} & --- & --- & 97.3$_{\scriptscriptstyle\pm0.5}$ & 97.0$_{\scriptscriptstyle\pm0.4}$ & 94.0$_{\scriptscriptstyle\pm0.6}$ \\
 & \cellcolor{zebracol}Tkun & \cellcolor{zebracol}--- & \cellcolor{zebracol}--- & \cellcolor{zebracol}\textbf{98.8$_{\scriptscriptstyle\pm0.2}$} & \cellcolor{zebracol}\textbf{98.3$_{\scriptscriptstyle\pm0.2}$} & \cellcolor{zebracol}\textbf{97.9$_{\scriptscriptstyle\pm0.3}$} \\
 & Bytedance & --- & --- & 97.9$_{\scriptscriptstyle\pm0.3}$ & 96.8$_{\scriptscriptstyle\pm0.4}$ & 95.0$_{\scriptscriptstyle\pm0.5}$ \\
 & \cellcolor{zebracol}Aria-AMT & \cellcolor{zebracol}--- & \cellcolor{zebracol}--- & \cellcolor{zebracol}\underline{98.3$_{\scriptscriptstyle\pm0.3}$} & \cellcolor{zebracol}\underline{97.6$_{\scriptscriptstyle\pm0.3}$} & \cellcolor{zebracol}\underline{96.4$_{\scriptscriptstyle\pm0.5}$} \\
 & MT3 & --- & --- & 95.6$_{\scriptscriptstyle\pm0.6}$ & 95.7$_{\scriptscriptstyle\pm0.4}$ & --- \\
\bottomrule
\end{tabular}
\vspace*{-\baselineskip}
\end{table}

\subsection{Temporal Grounding}
\label{sec:related}
We further evaluate Rubato on beat tracking and MIDI note detection (conventional AMT). We report the F1 score for both tasks using 70\,ms and 50\,ms tolerances respectively, following the standard conventions for each task.\footnote{For MIDI grounding we capture only the note onset; timestamp-based offset conventions and their downstream effects are discussed in \autoref{app:midi_with_offset}.} We compare with Beat-This~\citep{foscarin2024beatthis}, a dedicated spectrogram-to-beat-activation model and the current state of the art in beat tracking and Tkun~\citep{yan2024scoring}, Aria-AMT~\citep{edwards2024aria_amt}, Bytedance~\citep{kong2021high}, and MT3~\citep{gardner2021mt3_amt}, dedicated AMT systems. We evaluate three different Rubato dialects (TAST, DBD, and AMT). DBD and AMT are task specific dialects; TAST enables evaluation of both tasks with two key differences. Unlike DBD, TAST encodes downbeats as barline tokens, and beat positions are inferred at evaluation time from the metrical structure. Additionally, where AMT transcribes notes as \textit{performed}, TAST transcribes them as \textit{written}.\footnote{For example, a trilled note is a written as a single score object but produces multiple MIDI events whose count and timing vary across performances; similarly, an arpeggiated chord is one score moment but a spread sequence of onsets in MIDI.}

Rubato (DBD) surpasses Beat-This on beat detection and is slightly better for downbeats within confidence intervals (\autoref{tab:grounding}). Notably, Rubato (TAST) achieves 2.6 points higher \emph{downbeat} F1 than Rubato (DBD), while inferred beat positions do not show the same improvement. This suggests that predicting downbeats as barline tokens jointly with score structure provides additional context: barline placement is coupled to measure grouping and notational consistency, exposing the downbeat signal to richer structural cues than beat annotations. Rubato (AMT) reaches 97.0 note F1 on MAESTRO (\autoref{tab:grounding}), 1.3 points below Tkun (98.3) and 1.3 points above MT3 (95.7), demonstrating that a unified model remains competitive with dedicated AMT systems. Rubato (TAST) reaches note F1 scores of 87.1 on MAESTRO and 91.0 on ASAP, reflecting a fundamental difference in target: AMT optimizes for reproducing performed events, while TAST prioritizes a consistent, score-aligned representation. The resulting gap in note F1 largely follows from score–performance mismatch rather than transcription error; see \autoref{app:midi_with_offset} for further discussion.

\section{Analysis}
\label{sec:analysis}

OMR-NED measures how closely a transcript matches a canonical engraving, but not whether it preserves performance-specific information. A transcription system faithful to both the composition and the performance should also preserve the artistic identity that distinguishes one rendition from another. ATEPP, with multiple performers recording the same works, lets us evaluate both properties---compositional identity and performer-specific variation---through transcript-based retrieval without retrieval-specific training. We compute two metrics using $n$-gram shingling~\citep{broder1997resemblance}. Because the full $n$-gram sweep and stability analysis are reported in \autoref{app:version_matching}, \autoref{tab:version_matching} summarizes the best observed operating point for each representation.

\textbf{Work identity} (MAP$_\text{work}$): given a transcript, can we retrieve other recordings of the same piece? \textbf{Performer identity} (MAP$_\text{performer}$): given a transcript, can we distinguish recordings by the same performer from those by others? We evaluate two transcript variants: timestamps stripped (\textbf{Rubato}) and timestamps converted to relative deltas (\textbf{+\emph{relative timestamps}}), where each absolute timestamp is replaced by the elapsed time since the previous timestamp. 

\subsection{Work Identity}
\label{sec:trans_as_ret}
A transcript naturally supports retrieval-based downstream tasks; just as ASR transcripts enable text-based content deduplication~\citep{radford2023whisper} and search~\citep{garofolo2000trec} over audio, music transcripts allow for content-based matching without audio dependency.
We evaluate whether transcript-based representations support retrieval by comparing against retrieval-trained audio models: CLEWS~\citep{serra2025clews} and CoverHunterC~\citep{liu2023coverhunter}.\footnote{Audio baselines use checkpoints (\texttt{dvi-clews}, \texttt{dvi-coverhunterc}) and inference code from the CLEWS repository~\citep{serra2025clews}.}

\autoref{tab:version_matching} shows that despite not being trained for retrieval, all transcription systems are competitive with the audio-based, retrieval-trained baselines. Every transcription system surpasses CoverHunterC on MAP$_\text{work}$, and Rubato even falls within the confidence interval of CLEWS --- still the best performing system on MAP$_\text{work}$.
Furthermore, \autoref{app:version_matching} demonstrates that Rubato consistently achieves the highest performance of transcription models across all n-gram sizes; agreement at larger $n$ reflects structural consistency over longer musical spans. We also compare conventional tokenizations for the baseline transcription models (see \emph{N-gram (external)} in \autoref{tab:version_matching} and \autoref{fig:n_gram_tokenizer_isolation} in \autoref{app:version_matching}). 
The same transcriptions perform worse when represented with conventional tokenizers compared to when they are canonicalized with \intermo{}, suggesting that canonicalization contributes to retrieval quality independently of the upstream transcription model.\looseness=-1

\subsection{Performer Identity}
\label{sec:version_matching}
While work identity tells us whether the model captures \emph{what} was played, performer identity tests whether it captures \emph{how}. A system that returns a memorized canonical score would produce near-identical transcripts across performances of the same piece, collapsing MAP$_\text{performer}$ toward chance. Performer identity---retrieving other recordings by the same performer from transcript similarity alone---requires the system to preserve expressive detail.

\autoref{tab:version_matching} reveals a tradeoff among conventional cascades. Tkun$\to$M2ST achieves strong MAP$_\text{work}$ but scores among the lowest on MAP$_\text{performer}$: its sheet music conversion process canonicalizes compositional content at the cost of discarding performer-specific timing and phrasing. Tkun$\to$PM2S shows the opposite pattern, preserving performer identity while its MAP$_\text{work}$ degrades at larger $n$ (\autoref{app:version_matching}), indicating that its transcriptions lack long-range structural coherence.

Rubato does not exhibit this tradeoff. Even without timestamps---evaluating only the portion rendered as sheet music---it surpasses Tkun$\to$M2ST on MAP$_\text{performer}$ by 10 points while achieving a lower (better) OMR-NED. When relative timestamps are included (Rubato~+\emph{relative timestamps}), MAP$_\text{performer}$ rises to 72.3\%, within the confidence interval of Tkun$\to$PM2S---which, however, suffers from considerably worse OMR-NED. No other system achieves competitive OMR-NED, MAP$_\text{work}$, and MAP$_\text{performer}$ simultaneously on ATEPP.

Score readability and performer-fingerprint preservation need not be in conflict: Rubato achieves both from a single transcript, aided by a representation that is canonical in content and monotonic in time. \intermo{} provides both properties, and its benefits extend beyond Rubato: all PM2S cascades achieve higher MAP$_\text{work}$ and MAP$_\text{performer}$ when their outputs are re-represented with \intermo{} rather than REMI, and M2ST cascades improve over their native compound tokenization. The canonicalization is a representational contribution independent of the model trained on it.

\definecolor{zebracol}{gray}{0.965}
\definecolor{sectioncol}{gray}{0.88}

\begin{table}[t]
\centering
\caption{Version matching:  retrieval of different recordings of the same work and performer from transcript similarity alone ($n$-gram Jaccard) or from learned audio embeddings. MAP ($\uparrow$, \%) evaluated on ATEPP (106 pieces, 1495 recordings). Results shown at the best $n^*$ per system per metric; $n$-gram sweep curves and stability analysis in \autoref{app:version_matching}.}
\label{tab:version_matching}
\small
\setlength{\tabcolsep}{5pt}
\begin{tabular}{p{1.8cm} l cc}
\toprule
 & & \multicolumn{1}{c}{MAP$_\text{work}$} & \multicolumn{1}{c}{MAP$_\text{performer}$} \\
\midrule
\multirow{5}{1.8cm}{\textit{N-gram (InterMo)}}
 & \textbf{Rubato (TAST)} & \underline{97.4$_{\scriptscriptstyle\pm0.5}$} & 59.3$_{\scriptscriptstyle\pm2.0}$ \\
 & \cellcolor{zebracol}\hspace{2em} + \textit{relative timestamps} & \cellcolor{zebracol}96.6$_{\scriptscriptstyle\pm0.6}$ & \cellcolor{zebracol}\underline{72.3$_{\scriptscriptstyle\pm1.9}$} \\
 & Beat-This $\to$ Piano-A2S & 85.7$_{\scriptscriptstyle\pm1.0}$ & 48.5$_{\scriptscriptstyle\pm2.0}$ \\
 & \cellcolor{zebracol}Tkun $\to$ M2ST & \cellcolor{zebracol}96.1$_{\scriptscriptstyle\pm0.6}$ & \cellcolor{zebracol}48.5$_{\scriptscriptstyle\pm2.0}$ \\
 & Tkun $\to$ PM2S & 95.5$_{\scriptscriptstyle\pm0.7}$ & \textbf{73.3$_{\scriptscriptstyle\pm1.8}$} \\
\midrule
\multirow{2}{1.8cm}{\textit{N-gram (external)}}
 & \cellcolor{zebracol}Tkun $\to$ M2ST (M2ST-tok) & \cellcolor{zebracol}87.4$_{\scriptscriptstyle\pm1.2}$ & \cellcolor{zebracol}43.5$_{\scriptscriptstyle\pm2.0}$ \\
 & Tkun $\to$ PM2S (REMI) & 90.5$_{\scriptscriptstyle\pm0.9}$ & 67.7$_{\scriptscriptstyle\pm1.9}$ \\
 \midrule
\multirow{2}{1.8cm}{\textit{Audio-based}}
 & CLEWS & \textbf{97.6$_{\scriptscriptstyle\pm0.5}$} & 67.8$_{\scriptscriptstyle\pm1.9}$ \\
 & \cellcolor{zebracol}CoverHunterC & \cellcolor{zebracol}70.1$_{\scriptscriptstyle\pm1.2}$ & \cellcolor{zebracol}55.8$_{\scriptscriptstyle\pm2.1}$ \\
\bottomrule
\end{tabular}
\end{table}
\section{Conclusion}
\label{sec:conclusion}

Rubato transcribes piano music into time-aligned scores, in which notation and temporal grounding are predicted jointly rather than recovered through staged conversion. It produces more accurate piano sheet music than cascaded systems built from strong specialized components, including oracle variants that receive ground-truth intermediate signals. These results suggest that improving intermediate note or metrical predictions alone does not guarantee better final score transcription; part of the difficulty lies in what intermediate representations preserve or discard before score reconstruction.

\intermo{} provides a notation-native, time-grounded representation that keeps score structure and temporal information within a single transcript, making joint prediction compatible with standard autoregressive sequence modeling. Beyond notation quality, the resulting transcripts support downstream analyses of work identity and performer-specific variation. More broadly, our results suggest that representation design can offer an effective alternative to staged pipelines for score transcription.

\begin{ack}
We thank William Chen, Gon\c{c}alo Faria, Alisa Liu, William Merrill, Yi\u{g}itcan \"Ozer, Luiza Pozzobon, Pedro Ramoneda, and Shinji Watanabe for valuable discussions and feedback on earlier versions of this work.
\end{ack}

\bibliographystyle{unsrtnat}
\bibliography{references}

\appendix

\section{Gemini Evaluation}
\label{app:gemini_prompts}

We evaluate Gemini~3.1~Pro to examine how a frontier audio-language model, without task-specific training, approaches piano score transcription when prompted to produce conventional notation formats.

\subsection{Output Format Selection}

While none of the conventional notation formats---MusicXML, Humdrum **kern, ABC---provide structural properties that would help a language model ground its output in the audio it hears or reason about musical structure, they differ greatly in verbosity and in the amount of corresponding web data available for training. In a pilot test on a simple Bach fugue from ASAP (where all specialist systems produce low OMR-NED), we tested all three formats with both Gemini~3.1~Pro and Gemini~2.5~Pro. Both models consistently failed on MusicXML: Gemini~2.5~Pro produced truncated XML (parseable but nearly empty, 45 of 8{,}472 ground-truth symbols), while Gemini~3.1~Pro's output could not be parsed at all. **kern fared slightly better with Gemini~3.1~Pro (282 symbols, OMR-NED 97.7) but failed entirely with 2.5~Pro due to invalid octave notation. ABC notation yielded the most complete output: Gemini~3.1~Pro generated 1{,}463 symbols at OMR-NED 94.97, while Gemini~2.5~Pro produced only 12 symbols (OMR-NED 99.86). We therefore used Gemini~3.1~Pro with ABC notation for all subsequent experiments.

\subsection{Prompts}

We evaluate three prompting strategies: Baseline, Reasoning, In-context learning (ICL).

\paragraph{Baseline.} The system prompt tells the model to directly output ABC, without any additional output:
\begin{quote}
\small\ttfamily
You transcribe piano performance audio into symbolic notation.  
Return only valid abc content for the provided audio. 
No markdown fences, no explanations, no comments, and no extra text.
\end{quote}
The user message contains only the audio file.

\paragraph{Reasoning.} The model is prompted to output plain text for reasoning purposes before generating the final ABC representation. We extract the content from the last fenced block.
\begin{quote}
\small\ttfamily
You transcribe piano performance audio into symbolic notation. 
Use chain-of-thought style reasoning in plain text when it helps improve accuracy.
Prefer returning the final answer as valid ABC inside one fenced markdown block that starts with \textasciigrave\textasciigrave\textasciigrave abc and ends with \textasciigrave\textasciigrave\textasciigrave
\end{quote}

\paragraph{In-context learning (ICL).} We use the same system prompt as the Baseline. Before the test audio, we provide several worked examples, each consisting of an audio file followed by its ground-truth ABC transcription, drawn from the ASAP training split. The user prompt for each example is: ``Example input audio. Produce only ABC notation.''

\subsection{Transcription as Recall}
  We evaluate on the 123 unique ASAP test recordings (102 from the standard
  split and 25 from the Beyer split, with 4 overlapping). Under baseline                          
  prompting, 14 returned empty outputs (12 standard, 2 Beyer), typically for
  the longest pieces (e.g., Liszt Transcendental \'Etudes, Beethoven Waldstein                    
  Sonata); these are excluded from OMR-NED. Across all prompting modes,                           
  OMR-NED remains between 97.6 and 98.9                                                           
  (\autoref{tab:notation_quality})---near the 100\% ceiling, indicating that                      
  the generated notation has little overlap with the reference score. Given                       
  this level of performance and the associated API costs, we did not evaluate
  Gemini on the larger ATEPP dataset.                                                             
                  
  Manual inspection of the chain-of-thought traces suggests that Gemini                           
  approaches transcription as a recognition-and-recall task rather than
  acoustic analysis: 72\% of reasoning-prompted predictions open by naming the                    
  piece and composer, then attempt to emit a memorized canonical score. This
  behavior is further quantifiable because Gemini voluntarily fills the title                     
  (\texttt{T:}) and composer (\texttt{C:}) header fields in its ABC output,
  which we match against the ground-truth catalogue across all prompting modes.                   
   
    Under baseline prompting and counting once per recording, 16 of 109
  non-empty predictions (14.7\%) correctly identify both composer and piece. A further 11
  name the correct composer but misidentify the piece, and 40 produce a                           
  plausible piece number but attribute it to the wrong composer; the remaining
  42 identify neither correctly. Among incorrect attributions, the model
  disproportionately defaults to Chopin: the baseline misattributes 29\% of                       
  non-Chopin pieces to Chopin (28/96; non-Chopin denominators vary across                         
  modes because each produces a different number of empty outputs), consistent
  with Chopin's over-representation in web piano corpora. Reasoning prompts appear to amplify
  this bias: Chopin misattribution rises to 42\% (42/100), suggesting that the
  chain-of-thought reinforces the model's prior rather than relying on
  acoustic cues. In-context learning partially corrects it: the
  misattribution rate drops to 14\% (15/107), and correct identification rises                    
  from 14.7\% to 20.3\% (25/123).                                                                
                  
  Despite these differences in identification accuracy, OMR-NED barely moves                      
  across prompting modes (baseline 98.6, ICL 97.6; lower is better). The
  modest improvement tracks correct piece identification---retrieving a more                      
  relevant memorized score---rather than improved transcription of the audio.
  This pattern---where a model defaults to recalling memorized scores rather                      
  than transcribing the performance it hears---motivated the analysis                             
  in~\autoref{sec:trans_as_ret}, where we examine the broader relationship
  between transcription and retrieval across all evaluated systems.  

\renewcommand{\topfraction}{.9}
\renewcommand{\bottomfraction}{.5}
\renewcommand{\textfraction}{.1}
\setcounter{topnumber}{3}
\setcounter{bottomnumber}{1}
\setcounter{totalnumber}{4}

\section{MIDI Note Detection}
\label{app:midi_with_offset}

This appendix examines two aspects of MIDI-level evaluation: \autoref{app:offset} analyzes the effect of offset conventions (KeyOff vs.\ PedOff) on downstream score transcription and retrieval, and \autoref{app:mismatch} quantifies how much of the note F1 gap between TAST and AMT reflects score--performance mismatch rather than transcription error.

\subsection{Offset Conventions}
\label{app:offset}

While the MIDI transcription literature agrees on Note-F1 as the semitone-level pitch + onset timestamp, what is called an \emph{offset} without score notation is a more nuanced concept. A Disklavier records the key-release time and the sustain-pedal state
(MIDI CC64, a continuous $0$--$127$ stream) independently. Each note
therefore admits two offsets: \textbf{KeyOff}, the moment the performer
lifts off the key, and \textbf{PedOff}, KeyOff extended forward to the
point where CC64 drops below~$64$. In legato passages with a held pedal the two
can differ by hundreds of milliseconds. Prior work
evaluates PedOff by default~\citep{hawthorne2018maestro}, which is assumed to correspond to ``perceived'' offset. Human studies verify that sound offset perception is more context-dependent than that of onsets~\citep{ali2025distinctive}, and we hypothesize that it depends more on tonal and metrical context~\citep{mehr2025core}---something that neither KeyOff nor PedOff can effectively capture from physical mechanisms alone.

Without an assumption about what is being perceived, predicting both the key-release time and the sustain-pedal state separately is still useful for recovering the sound-production mechanism from audio. In particular, being able to recover KeyOff---the offset associated with the performer's fingering movement---has clear relevance for downstream tasks such as expressive performance modelling and generation, multimodal music analysis including hand movements, and computer-assisted music education. To
recover these performance gestures from audio, Rubato (when prompted with the AMT task) emits note offsets and pedal events as separate tokens, predicting the key-release time and the CC64 state
independently.

\begin{figure}[t]
    \centering
    \includegraphics[width=\textwidth]{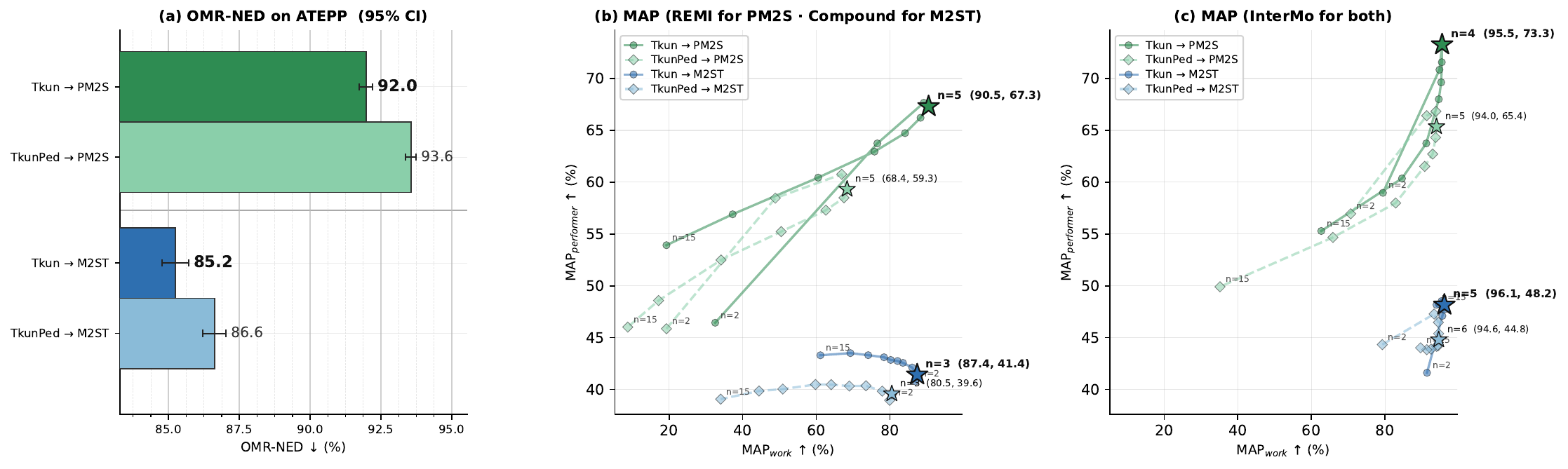}
    \caption{Downstream impact of the upstream offset convention.
    Only the upstream AMT changes between Tkun (KeyOff finetune) and
    TkunPed (PedOff training); every downstream model is held fixed.
    (a) OMR-NED on ATEPP score transcription via the PM2S and M2ST
    cascades (lower is better). (b) Retrieval MAP with REMI and
    Compound tokenizers. (c) Retrieval MAP with \intermo{}. Tkun (when switched to predicting performer's key offset instead of the conventional `pedal offset' rule)
    feeds better MIDI into every downstream system.}
    \label{fig:midi_offset_downstream}
\end{figure}

\citet{yan2024scoring} first raised the issue of offset conventions by publishing their
paper using the PedOff convention (TkunPed; named \texttt{Transkun-V2}\footnote{\url{https://github.com/Yujia-Yan/Transkun/tree/main}, see Model Cards.} in their model card) and subsequently releasing a
KeyOff-finetuned variant (Tkun; named \texttt{Transkun-V2-NoExt}, \emph{no pedal extension}, in their model card), noting that key-release offsets are more
realistic for downstream use.
\autoref{fig:midi_offset_downstream} evaluates this claim end-to-end
by swapping \emph{only} the upstream AMT between the two checkpoints
while every downstream model remains untouched. On ATEPP score
transcription, OMR-NED improves by $\sim$$1.5$~pp on both the PM2S and
M2ST cascades (panel a: Tkun$\to$PM2S $92.0$ vs.\ TkunPed$\to$PM2S
$93.6$; Tkun$\to$M2ST $85.2$ vs.\ TkunPed$\to$M2ST $86.6$). On
retrieval, both MAP@work and MAP@performer improve, and this holds
across tokenizers---REMI (a score MIDI tokenizer, used for PM2S) and Compound (the tokenizer used in M2ST)  (panel b) as well as
\intermo{} (panel c). We therefore believe that KeyOff is not only more meaningful for physical performance modelling, but also for downstream uses, as exemplified by score notation quality and retrieval performance.

\begin{figure}[t]
    \centering
    \includegraphics[width=\textwidth]{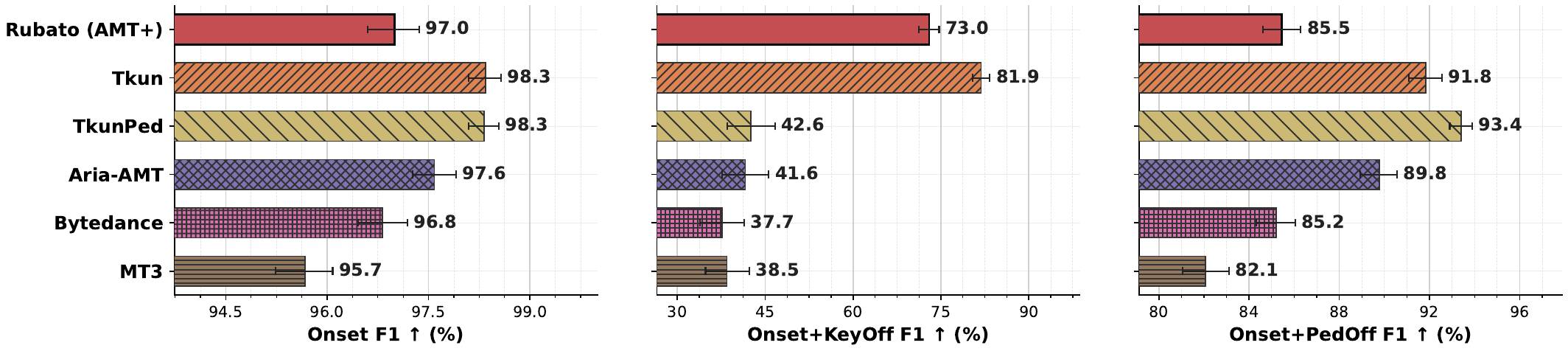}
    \caption{MIDI note-detection F1 on MAESTRO test (177 files).
    TkunPed is Transkun-V2 as published (PedOff training); Tkun is the
    authors' subsequent KeyOff finetune of the same model.}
    \label{fig:midi_offset_metrics}
\end{figure}

\makeatletter
\renewcommand{\fps@table}{t!}
\makeatother
\begin{table}[t]
  \caption{MIDI transcription F1 metrics on the MAESTRO test set (177 files), comparing the KeyOff and PedOff offset definitions. Values are mean \%~$\pm$~half-95\,\% CI (bootstrap, $n=10{,}000$). \textbf{Bold}: best per column. N/A: model does not predict velocity.}

  \centering
  \small
  \begin{tabular}{lrrrrrr}
  \toprule
  \textbf{Model} & \shortstack{\small Note\\F1} & \shortstack{\small N+Vel\\F1} & \shortstack{\small N+KeyOff\\F1} & \shortstack{\small N+KeyOff+Vel\\F1} & \shortstack{\small N+PedOff\\F1} & \shortstack{\small N+PedOff+Vel\\F1} \\
  \midrule
  Rubato (AMT) & 97.0$_{\scriptscriptstyle\pm0.4}$ & 94.0$_{\scriptscriptstyle\pm0.6}$ & 73.0$_{\scriptscriptstyle\pm1.8}$ & 70.7$_{\scriptscriptstyle\pm1.8}$ & 85.5$_{\scriptscriptstyle\pm0.8}$ & 82.8$_{\scriptscriptstyle\pm0.9}$ \\
  \cellcolor{zebracol}Tkun & \cellcolor{zebracol}\textbf{98.3$_{\scriptscriptstyle\pm0.2}$} & \cellcolor{zebracol}\textbf{97.9$_{\scriptscriptstyle\pm0.3}$} & \cellcolor{zebracol}\textbf{81.9$_{\scriptscriptstyle\pm1.5}$} & \cellcolor{zebracol}\textbf{81.5$_{\scriptscriptstyle\pm1.5}$} & \cellcolor{zebracol}91.8$_{\scriptscriptstyle\pm0.7}$ & \cellcolor{zebracol}91.4$_{\scriptscriptstyle\pm0.8}$ \\
  TkunPed & \textbf{98.3$_{\scriptscriptstyle\pm0.2}$} & 97.7$_{\scriptscriptstyle\pm0.3}$ & 42.6$_{\scriptscriptstyle\pm4.1}$ & 42.5$_{\scriptscriptstyle\pm4.1}$ & \textbf{93.4$_{\scriptscriptstyle\pm0.5}$} & \textbf{92.9$_{\scriptscriptstyle\pm0.5}$} \\
  \cellcolor{zebracol}Aria-AMT & \cellcolor{zebracol}97.6$_{\scriptscriptstyle\pm0.3}$ & \cellcolor{zebracol}96.4$_{\scriptscriptstyle\pm0.5}$ & \cellcolor{zebracol}41.6$_{\scriptscriptstyle\pm4.0}$ & \cellcolor{zebracol}41.3$_{\scriptscriptstyle\pm4.1}$ & \cellcolor{zebracol}89.8$_{\scriptscriptstyle\pm0.8}$ & \cellcolor{zebracol}88.8$_{\scriptscriptstyle\pm0.9}$ \\
  Bytedance & 96.8$_{\scriptscriptstyle\pm0.4}$ & 95.0$_{\scriptscriptstyle\pm0.5}$ & 37.7$_{\scriptscriptstyle\pm3.8}$ & 37.3$_{\scriptscriptstyle\pm3.8}$ & 85.2$_{\scriptscriptstyle\pm0.9}$ & 83.7$_{\scriptscriptstyle\pm0.9}$ \\
  \cellcolor{zebracol}MT3 & \cellcolor{zebracol}95.7$_{\scriptscriptstyle\pm0.4}$ & \cellcolor{zebracol}N/A & \cellcolor{zebracol}38.5$_{\scriptscriptstyle\pm3.7}$ & \cellcolor{zebracol}N/A & \cellcolor{zebracol}82.1$_{\scriptscriptstyle\pm1.0}$ & \cellcolor{zebracol}N/A \\
  \bottomrule
\end{tabular}
\label{tab:midi_offset_metrics}
\end{table}

\makeatletter
\renewcommand{\fps@table}{tbp}
\makeatother

\autoref{tab:midi_offset_metrics} and
\autoref{fig:midi_offset_metrics} report \texttt{mir\_eval} F1 on the
MAESTRO test split (177 files) under both conventions, explaining the
source of this downstream gap. Prior AMT models predict PedOff directly, so evaluation is inherently asymmetric: ground-truth PedOff is constructed by extending GT KeyOff with GT CC64, while the model side is taken as-is with no corresponding decomposition. When a model predicts KeyOff and CC64 as separate streams---as Rubato and Tkun (V2-NoExt) do---the same extension convention can be applied to both sides: GT KeyOff extended by GT CC64, predicted KeyOff extended by predicted CC64. This symmetric setup lets the KeyOff metric isolate key-release accuracy from pedal prediction, and the PedOff metric evaluate both components jointly. Note F1 and Note+Vel F1 are
offset-independent and agree closely across systems. The two Transkun
checkpoints differ by $\sim$$40$ points on KeyOff F1 ($81.9$ vs.\
$42.6$) and swap by only $1.6$ points on PedOff F1 ($91.8$ vs.\
$93.4$), so a downstream model consuming KeyOff-style MIDI receives a
very different signal depending on which upstream sits in front of
it. AMT-prompted Rubato reaches $73.0$ KeyOff F1, below Tkun's specialist finetune but above the other models.

\begin{figure}[t]
    \centering
    \includegraphics[width=0.36\textwidth]{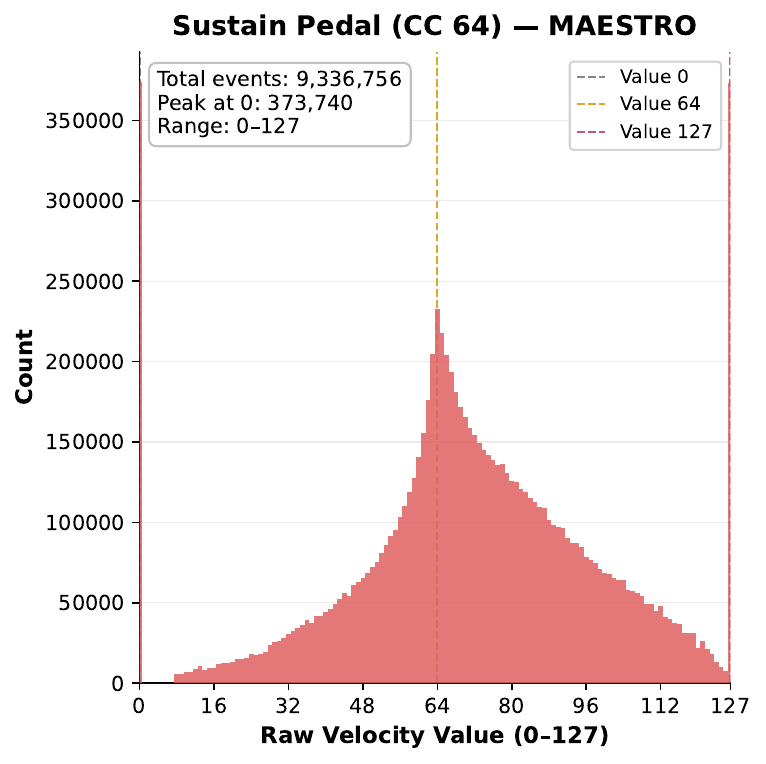}
    \caption{Raw sustain-pedal velocity distribution over $9.3$M MAESTRO sustain-pedal events.
    Besides the expected peaks at $0$ and $127$, a sharp local peak
    sits \emph{exactly on} the binarisation threshold of $64$, so the
    on/off decision is most ambiguous precisely where it is most often
    made.}
    \label{fig:pedal_hist}
\end{figure}

\autoref{fig:pedal_hist} plots the raw CC64 distribution across the
$9.3$M sustain events in MAESTRO. The peaks at $0$ ($\sim$374k,
released) and $127$ ($\sim$373k, fully depressed) are expected; the
interior distribution, however, has a sharp local peak \emph{exactly at
$64$} ($\sim$233k, the global maximum of the interior), the value at
which CC64 is binarised. A hard threshold at this point is maximally
sensitive: small perturbations flip the binary pedal state and shift
reported offsets by hundreds of milliseconds, so an evaluation that
depends on this threshold is, in part, evaluating the threshold itself.
We recommend that AMT evaluations state which offset convention they
report, extend predictions with their own pedal events when reporting
PedOff, and prefer KeyOff where the downstream task permits---on the
same finetuning base, switching from KeyOff to PedOff moves
onset+offset F1 from $42.6$ to $93.4$, a $50.8$-point swing that is
almost entirely an artefact of the convention.

\subsection{Target Mismatch Quantification}
\label{app:mismatch}
To quantify how much of the gap in Rubato (TAST) note F1 performance described in \autoref{sec:related} reflects model error versus target mismatch, we analyze the note-level score$\leftrightarrow$performance alignments in n-ASAP~\citep{peter2023n_asap}. Even perfectly aligned score annotations achieve only 93.4 note F1 under this metric. The largest source of discrepancy (4.4 F1 points) arises from genuine performer–score differences, with 4.9\% of score notes and 4.3\% of performed notes having no counterpart in the alignment. An additional 2.2 points is attributable to collapsing arpeggiated or spread events into single score moments.

This distinction becomes clearer when comparing against cascaded systems on ASAP. When prompted for AMT, Rubato achieves higher note F1 than Tkun$\to$PM2S (97.3 vs.\ 95.2), but does not explicitly model score structure. In contrast, TAST trades note-level accuracy (91.0 note F1) for substantially improved structural grounding: it achieves 45 points higher downbeat F1 (67.8 vs.\ 22.4) and 20 points higher beat F1 (75.8 vs.\ 56.0). This advantage holds despite TAST emitting only barlines explicitly and inferring intra-bar beats post hoc.

\section{Version Matching and the Effect of Long-Range N-grams}
\label{app:version_matching}

In document matching, long-range n-grams are used for finding exact duplicates. Retrieval at n=2,3 may be explained by just local surface-level relationships being preserved across versions, but retrieval at long-range n-grams requires verbatim agreement. In music, short n-grams capture note-to-note transitions; longer spans require at least rhythmic structure, metrical grouping, and voice-leading consistency to match. How a system's MAP degrades as n grows tells us whether its transcriptions share only local fragments across versions or preserve larger-scale structure.

\begin{figure}[t]
    \centering
    \includegraphics[width=\linewidth]{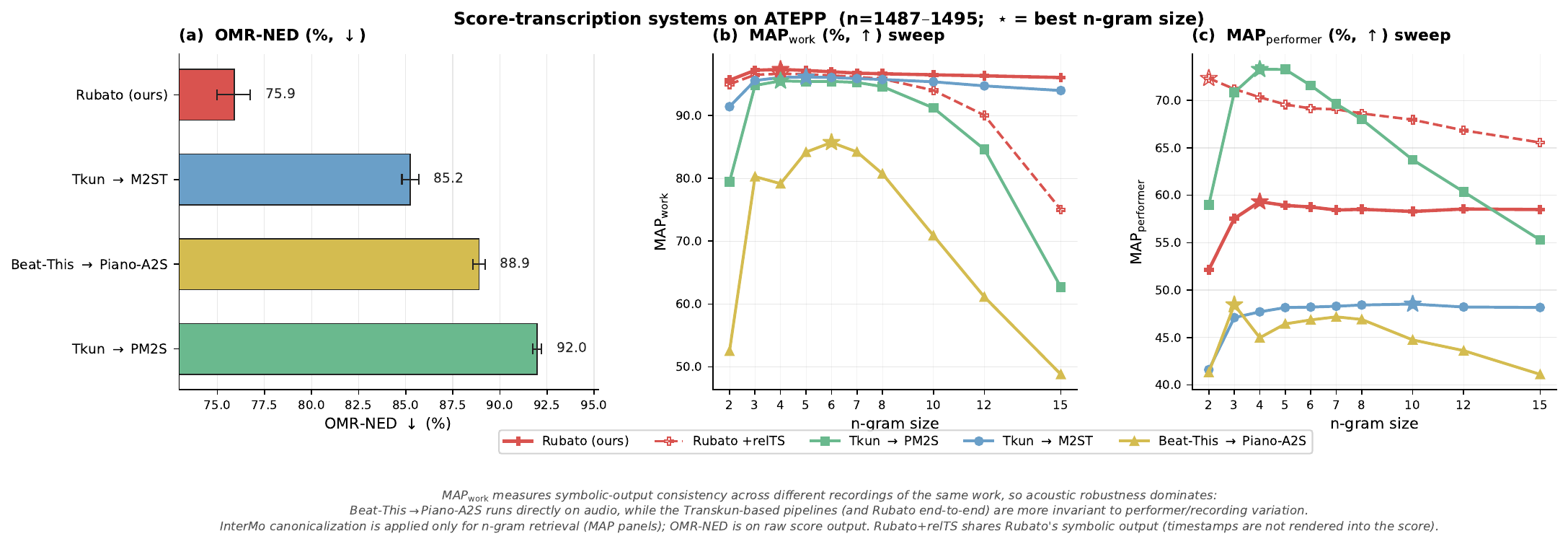}
    \caption{Score-transcription systems on ATEPP (All represented with InterMo). (a)~OMR-NED on raw score output. (b,c)~MAP$_\text{work}$ and MAP$_\text{performer}$ as a function of n-gram size, with InterMo canonicalization applied for n-gram retrieval. Stars mark the best n-gram size per system. Rubato~+relTS shares Rubato's symbolic output; timestamps are not rendered into the score. Agreement at larger n requires long-range structural consistency across different versions and is hence more desirable.}
    \label{fig:n_gram_map_system_comparison}
\end{figure}

\subsection{System comparison (\autoref{fig:n_gram_map_system_comparison}).} 

\label{app_sec:n_gram_map_system_comparison}
Tkun$\to$M2ST and Tkun$\to$PM2S illustrate a tradeoff between structural consistency and performer preservation. M2ST remains stable on MAP$_\text{work}$ across n-gram sizes (panel~b) and has lower OMR-NED (panel~a), but scores among the lowest on MAP$_\text{performer}$ (panel~c): its score-level structuring canonicalizes compositional content at the cost of discarding performer-specific detail. PM2S shows the opposite pattern---it preserves more performer identity but its MAP$_\text{work}$ degrades at higher n, indicating that its transcriptions lack long-range structural coherence. Beat-This$\to$Piano-A2S follows a similar degradation pattern on MAP$_\text{work}$, and is particularly unstable across n-gram sizes because its audio encoder is not robust to real-world recording conditions in ATEPP (sourced from YouTube, with some recordings having heavy noise), whereas Transkun and Rubato do not suffer from the same issue.

\autoref{fig:n_gram_map_system_comparison} also reveals that the aforementioned work-performer tradeoff does not affect Rubato. It achieves the lowest OMR-NED and consistently maintains the best MAP$_\text{work}$ across n-gram sizes. On MAP$_\text{performer}$, even without timestamps (the portion rendered into sheet music), Rubato surpasses Tkun$\to$M2ST---its main competitor on OMR-NED---by 10 points, meaning the rendered sheet music is both closer to the reference and encodes more performer expressivity in the visual output without the same tradeoff. When relative timestamps are included (Rubato~+relTS), MAP$_\text{performer}$ rises to 72.3\%, within the confidence interval of Tkun$\to$PM2S, which has a considerably worse OMR-NED of 92.0. Rubato~+relTS also does not degrade heavily on MAP$_\text{performer}$ at larger n. This suggests that relative timestamps encode performer-specific patterns beyond local timing---potentially phrase-level interpretive signatures. We did not investigate what drives this long-range performer identity, but the observation may be worth further study.

\begin{figure}[t]
    \centering
    \includegraphics[width=\linewidth]{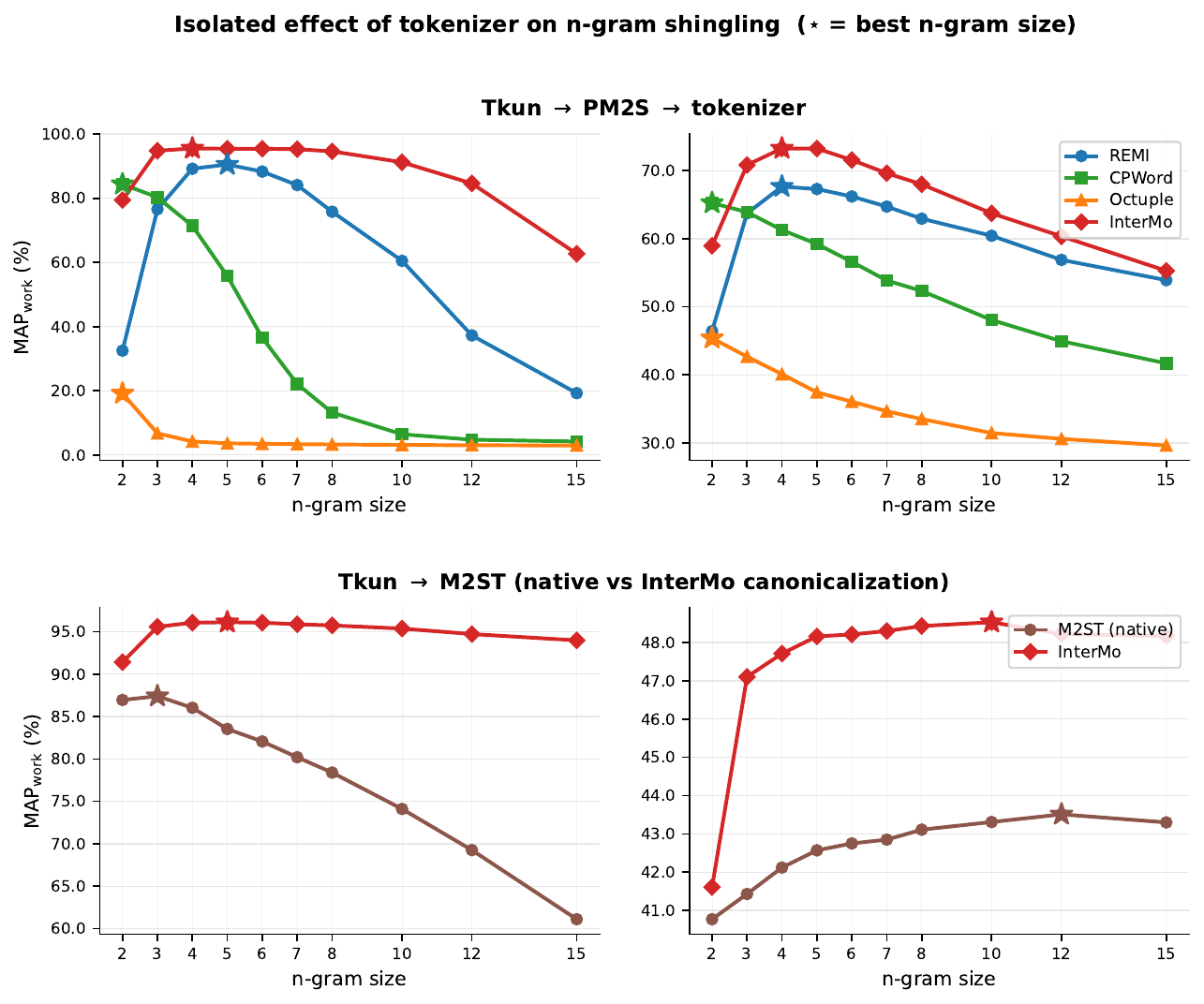}
    \caption{Isolated effect of tokenizer on n-gram shingling,
    holding the upstream transcription model (Transkun) fixed.
    Top: PM2S output tokenized with REMI, CPWord, Octuple, and
    InterMo. Bottom: M2ST native compound tokens vs.\ InterMo
    canonicalization. Left columns show MAP$_\text{work}$, right
    columns show MAP$_\text{performer}$. Stars mark the best
    n-gram size per tokenizer.}
    \label{fig:n_gram_tokenizer_isolation}
\end{figure}

\subsection{Tokenizer isolation (\autoref{fig:n_gram_tokenizer_isolation}).}
\label{app_sec:n_gram_tokenizer_isolation}
Tkun$\to$PM2S output is a \emph{score MIDI}, produced as a merger of five individual heads rather than a single token stream. To compare it reliably in this experiment, we tried different tokenizers from the MidiTok library and found that the best tokenizer for this retrieval scenario was REMI. The top portion of \autoref{fig:n_gram_tokenizer_isolation} holds Transkun's piano-roll output fixed and varies only the downstream tokenizer, isolating the effect of tokenization on n-gram consistency. On the PM2S cascade, InterMo maintains stable MAP$_\text{work}$ across n-gram sizes. REMI, CPWord, and Octuple all degrade---Octuple collapses toward near-zero beyond n$\approx$10, and CPWord drops substantially. On the M2ST cascade, the same pattern holds: InterMo sustains MAP$_\text{work}$ where M2ST's native compound tokens degrade. This isolates n-gram stability as a property of the representation, not the upstream model.

N-gram stability, as a representational property, is directly related to subword tokenization efficiency and learnability in full autoregressive architectures such as RNNs. While transformer decoders such as the one in Rubato do not technically require such n-gram stability, it universally reduces the permutation invariance in the next-token prediction objective. How musical events are grouped into tokens directly affects whether structural relationships survive across transcriptions of different performances. We argue that n-gram stability under varying canonicalizations of polyphony deserves further investigation, particularly in connection to tokenizer-free approaches and sub-word merging strategies. In this work we only explored what we call \emph{Interval-Piece}, InterMo pretokenized at score moments, since our aim was to insert expressivity-preserving timestamps in the generated sheet music. However, the long substring matching we observed suggests that relaxing the pretokenization constraints (e.g., a \emph{Bar-Piece} tokenization of InterMo) will allow better token efficiencies for purely symbolic generation. We hope future work explores the relationship between transcript n-grams, tokenization design, and music version matching on other InterMo tasks that do not require precise timestamps.

\end{document}